\documentclass[aps,prl,twocolumn,superscriptaddress,notitlepage,nofootinbib,longbibliography]{revtex4-2}
\usepackage{mathrsfs}
\usepackage{epsfig}
\usepackage{graphicx}
\usepackage{amsfonts}
\usepackage[figuresright]{rotating}
\usepackage{amssymb}
\usepackage{amsmath}
\usepackage{dcolumn}
\usepackage{bm}
\usepackage{comment}
\usepackage{xcolor}
\usepackage{color}
\usepackage{braket}
\usepackage{units}
\usepackage{xspace}

\usepackage{bbold}
\definecolor{mypink3}{cmyk}{0, 0.7808, 0.4429, 0.1412}
\definecolor{mypink1}{rgb}{0.858, 0.188, 0.478}
\definecolor{mypink2}{RGB}{219, 48, 122}
\usepackage[colorlinks=true, allcolors=mypink1]{hyperref}
\usepackage[fleqn]{mathtools}

\usepackage[shortlabels]{enumitem}
\newcommand{\ECNU}{Quantum Institute for Light and Atoms, State Key Laboratory of Precision Spectroscopy, Department of Physics, School of Physics and Electronic Science, East China Normal University, Shanghai 200062, China}
\newcommand{\SBH}{Shanghai Branch, Hefei National Laboratory, Shanghai 201315, China}
\newcommand{\SPA}{School of Physics and Astronomy, and Tsung-Dao Lee Institute, Shanghai Jiao Tong University, Shanghai 200240, China}
\newcommand{\SRC}{Shanghai Research Center for Quantum Sciences, Shanghai 201315, China}
\newcommand{\CIC}{Collaborative Innovation Center of Extreme Optics, Shanxi University, Taiyuan, Shanxi 030006, China}
\begin{document}
	\title{Enhanced Quantum Metrology with Non-Phase-Covariant Noise}
	
	\author{Jia-Xin Peng}
	\affiliation{\ECNU}
	\author{Baiqiang Zhu}
	\affiliation{\ECNU}
	\author{Weiping Zhang}
	\email{wpz@sjtu.edu.cn}
		\affiliation{\SPA}
	\affiliation{\SBH}
	\affiliation{\SRC}
	\affiliation{\CIC}
	\author{Keye Zhang}
	\email{kyzhang@phy.ecnu.edu.cn}
	\affiliation{\ECNU}
	\affiliation{\SBH}
	\begin{abstract}
		The detrimental impact of noise on sensing performance in quantum metrology has been widely recognized by researchers in the field. However, there are no explicit fundamental laws of physics stating that noise invariably weakens quantum metrology. We reveal that phase-covariant (PC) noise either degrades or remains neutral to sensing precision, whereas non-phase-covariant (NPC) noise can potentially enhance parameter estimation, surpassing even the ultimate precision limit achievable in the absence of noise. This implies that a non-Hermitian quantum sensor may outperform its Hermitian counterpart in terms of sensing performance. To illustrate and validate our theory, we present several paradigmatic examples of magnetic field metrology.
	\end{abstract}
	\maketitle
	\emph{Introduction.}---Investigating quantum parameter estimation in open systems is essential due to the unavoidable interaction of real physical systems with their surrounding environment \cite{escher2011quantum,haase2016precision}. Previous studies consistently show that environmental noise degrades quantum coherence, leading to reduced sensing precision. Strategies such as dynamic decoupling \cite{maze2008nanoscale,taylor2008high, DELL, Sekatski}, time optimization \cite{SALSS, CAW}, quantum error correction \cite{ZSSS, KLSC, HLKSK}, feedback control \cite{hirose2016coherent, LIUJJ, ZQWZ}, quantum trajectory monitoring \cite{MOLA} and Floquet engineering \cite{AJH} have been developed to overcome this challenge. Some works explore noise types with lesser detrimental effects, revealing that non-Markovian noises \cite{LXMPRA,ZNCW,ZZZWY} or noises with special orientation \cite{NMBY} can be advantageous.
	
	In fact, there is no fundamental law that prohibits the positive influence of environmental noise on quantum metrology. Recent findings recognize noise as a booster for quantum precision measurement and sensing in some cases \cite{EAPMS,PhysRevEM,TQSWXK,KLHCSQ}. For instance, a high-temperature reservoir can enhance system fluctuations, improving distinguishability in measuring dual electron spin states \cite{GYJXJS}. The theory of enhanced sensor sensitivity at environmentally induced exceptional points has been experimentally validated \cite{chen2017exceptional,kononchuk2022exceptional,hodaei2017enhanced}, emphasizing the use of environmental factors to amplify quantum sensor responses to weak signals. Moreover, a dissipative adiabatic measurement based on noise is proposed \cite{ZDJGJB}, where noise is an indispensable resource.

	Spirited by the development of noisy quantum metrology, two important questions naturally arise: (1) What types of noise may boost quantum metrology; (2) Can estimation precision in the presence of noise surpass the noiseless precision limit? This Letter aims to address these two questions. Firstly, we demonstrate that only NPC noise is likely to boost quantum metrology, while PC noise has a negative effect (or no effect) on sensing performance. Surprisingly, we find that the sensing precision of a non-Hermitian sensor influenced by NPC noise may surpass the ultimate precision limit given by its Hermitian counterpart. We illustrate these findings in the analysis of paradigmatic quantum metrological schemes, including quantum estimation of magnetic field strength and its direction. We emphasize that, unless otherwise stated, the noise mentioned below does not contain estimated parameters.

	\emph{Preliminaries.}---The dynamic evolution of an open quantum system is described by the master equation \cite{rivas2012open,breuer2002theory} $\partial _{t}\hat{\rho}%
	\left( t\right) =(\hat{\mathcal{H}}+\hat{\mathcal{L}})\left[ \hat{\rho}%
	\left( t\right) \right] $ (hereafter $\hbar =1$), where $\hat{\mathcal{H}}\left[ \bullet \right] :=-i[\hat{H},\bullet ]$ and $\hat{\mathcal{L}}\left[ \bullet \right] :=\sum_{k}\gamma _{k}[\hat{\Gamma}_{k}\bullet \hat{\Gamma}_{k}^{\dagger }-\frac{1}{2}\{\hat{\Gamma}_{k}^{\dagger }%
	\hat{\Gamma}_{k},\bullet \}]$, with $\hat{H}$ the
	Hamiltonian of system and $\hat{\Gamma}_{k}$ the quantum jump operator associated
	with a dissipative channel occurring at decay rate $\gamma _{k}$. 
	The series solution of $\hat{\rho}\left( t\right) $ reads as~\cite{RTSKX}
	\begin{equation}
	\label{Eq00}
	\hat{\rho}\left( t\right) =e^{\mathcal{\hat{H}} t}[\hat{\Pi}\left( t\right) [\hat{\rho}\left( 0\right) ]],
	\end{equation}%
	where $\hat{\rho} \left( 0\right) $ is
	the initial-state density matrix operator of system, $\hat{\Pi}\left( t\right)=\sum_{n=1}^{\infty }(\hat{\mathbb{1}}+%
	\hat{\Xi}_{n})$ is an effective dissipative superoperator, $\hat{\mathbb{1}}$ is identity superoperator, and
	\begin{widetext}
		\begin{equation}
		\label{Eq1}
		\hat{\Xi}_{n}=\int_{0}^{t}%
		\int_{0}^{t_{1}}\cdots \int_{0}^{t_{n-1}}e^{-\mathcal{\hat{H}}t_{1}}\hat{\mathcal{L}}%
		e^{\mathcal{\hat{H}}t_{1}}e^{-\mathcal{\hat{H}}t_{2}}\hat{\mathcal{L}}e^{\mathcal{\hat{%
					H}}t_{2}}\cdots e^{-\mathcal{\hat{H}}t_{n}}\hat{\mathcal{L}}e^{\mathcal{\hat{H}}%
			t_{n}}d{t_{n}}d{t_{n-1}}\cdots d{t_{1}}(n>0),
		\end{equation}%
	\end{widetext}
	with the integral upper limit satisfying $t>t_{1}>\cdots>t_{n-1}$.
	Based on the commutativity of noise-induced operations with Hamiltonian dynamics in the master equation, noise is categorized into two types. The first one is PC noise, where dissipative dynamics commutes with coherent dynamics \cite{PCNPC}, i.e,  $\mathcal{\hat{H}}[\mathcal{\hat{L}}[ \hat{\rho}( t)] ]=\mathcal{\hat{L}}[\mathcal{\hat{H}} [ 
	\hat{\rho}\left( t\right)] ]$. In this case $\hat{\rho}\left( t\right)$ can be expressed as~\cite{RTSKX},
	\begin{equation}
	\label{Eq2}
	\hat{\rho}\left( t\right) =e^{\mathcal{\hat{L}}t}[e^{\mathcal{\hat{H}}t}%
	[\hat{\rho}\left( 0\right)]] =e^{\mathcal{\hat{L}}t}[\hat{\rho}^{\left(
		0\right) }\left( t\right)] ,
	\end{equation}%
	where $\hat{\rho}^{\left( 0\right) }\left( t\right) $ represents the evolved state in the noiseless case. This implies a complete separation of the two dynamics in time evolution, and their order does not affect the final state $\hat{\rho}\left( t\right) $.
	
	The other is NPC noise, where the two dynamics are no longer commutative \cite{PCNPC}, i.e., $\mathcal{\hat{H}}[\mathcal{\hat{L}} [\hat{\rho} ( t ) ] ]\neq \mathcal{\hat{L}}[\mathcal{\hat{H}} [ \hat{\rho} ( t ) ] ]$. In this case, the two dynamics cannot be separated in state evolution, presenting significant challenges for solving the master equation.
	But at the short-term limit, $\hat{\rho}\left( t\right)$ is approximated as~\cite{RTSKX}
	\begin{equation}
	\label{Eq3}
	\hat{\rho}\left( t\right) \approx e^{\mathcal{\hat{H}}t}[e^{\mathcal{ \hat{L} } t}[\hat{\rho}\left( 0\right)]].
	\end{equation}
	The approximate expression involves transforming the two concurrent dynamics processes into a sequential order, with dissipative dynamics preceding the coherent dynamics. This sequence implies that the noise may solely alter the effective initial state which subsequently undergoes coherent evolution.
	
	Let $\theta $ represents the estimated parameter,  and the corresponding estimation error is quantified by quantum Cram\'er-Rao bound (QCRB) \cite{helstrom1969quantum}, i.e., Var$(\hat{\theta})\geq 1/\nu F_{\theta }$. 
	Here, Var$(\hat{\theta})$ is the mean squared error of unbiased estimator, $\nu$ is the number of trials, $F_{\theta}[\hat{\rho}_{\theta }]:=\text{Tr}[\hat{\rho}_{\theta }\hat{L}_{\theta }^{2}]$ is quantum Fisher information (QFI), and $\hat{L}_{\theta }$ is the symmetric logarithmic derivative formally defined by $\partial _{\theta }\hat{\rho}_{\theta } =(\hat{\rho}_{\theta } \hat{L}_{\theta }+\hat{L}_{\theta }\hat{\rho}_{\theta })/2$. The QCRB indicates the larger the QFI, the higher the theoretically achievable estimation precision of the sensor.

	\emph{Non-phase-covariant noise enhanced sensing performance.}---Assuming $\theta$ is only included in the Hamiltonian, superoperator $\hat{\mathcal{H}}\rightarrow\hat{\mathcal{H}}(\theta)$.
	Liouvillian superoperator $\mathcal{\hat{L}}$ is negative
	semidefinite when $\gamma _{k}\ \geq 0$ for $\forall k$  \cite{rivas2012open}. Its eigenvalues $\zeta _{j}$ and right (left) eigenmatrices $\hat{\Re}^{R}_{j}$ ($\hat{\Re}^{L}_{j}$) satisfy eigenequation  $\mathcal{\hat{L}}\hat{\Re}^{R}_{j}=\zeta _{j}\hat{\Re}^{R}_{j}$ ($\mathcal{\hat{L}}^{\dagger}\hat{\Re}^{L}_{j}=\zeta^* _{j}\hat{\Re}^{L}_{j}$) where $\mathbb{R}$e$\left[ \zeta _{j}\right] \leq 0$ for $\forall j$ ($j=1\sim d^{2}$, here $d$ is the dimension of the system), leading to $0\leq e^{\mathbb{R}\text{e}\left[ \zeta _{j}\right] }\leq 1$ and a potential decrease in the elements of the density matrix and information of estimated parameter. Thus, for PC noise, based on  Eq.~(\ref{Eq2}) and  the definition of the QFI we can conclude that \cite{FBDZM}
	\begin{equation}
	\label{Eq6}
	F_{\theta }\left[ \hat{\rho}_{\theta }\left(t\right) \right] \leq
	F_{\theta }[\hat{\rho}_{\theta }^{\left( 0\right) }\left(t\right) ].
	\end{equation}%
	The formula indicates that PC noise is detrimental to estimation precision, or at best, it does not affect it.
	This is because if information about $%
	\theta $ in $\hat{\rho}_{\theta }^{\left( 0\right) }\left(t\right) $ is not encoded in a decoherence-free subspace, it leaks to the environment, resulting in a decrease in estimation precision.
	
	For NPC noise, a positive answer to question 1 can be obtained by studying the following limiting scenario through Eq.~(\ref{Eq3}).
	Suppose initial state $\hat{\rho}\left(0\right) $ is an eigenstate of Hamiltonian $\hat{H}\left( \theta \right) $, without $e^{\mathcal{\hat{L}}t}$ we can't extract any information about $\theta $ from state $\hat{\rho} (t)$ since it only manifests as a global phase factor. However, introducing NPC noise causes $\hat{\rho}(0)$ to deviate from the eigenstate, and $\theta$ is subsequently encoded into $\hat{\rho} (t)$ under the action of $e^{\mathcal{\hat{H}}\left( \theta \right) t}$.
	This results in $F_{\theta}\left[ \hat{\rho}_{\theta}\left(t\right) \right] \neq 0$, signifying that NPC noise enables previously unattainable quantum parameter estimation.
	Furthermore, if $\hat{\rho}\left( 0\right)$ is not the optimal initial state, the action of $e^{\mathcal{\hat{L}}t}$ may bring $\hat{\rho}\left( 0\right)$ closer to the optimal state, leading to enhanced estimation precision.
	
	Now, addressing question 2: Can the sensing precision of a non-Hermitian sensor with NPC noise surpass its Hermitian counterpart's limit? If the noise itself includes the estimated parameter, a positive answer is not surprising, as recent research has also confirmed \cite{chen2020fluctuation}. This Letter focuses on the case where the noise lacks the estimated parameter.
	Unfortunately, this situation cannot be analyzed solely from Eq.~(\ref{Eq3}). This is because, under the encoding by the Hamiltonian, the performance of the effective initial state $e^{\mathcal{\hat{L}}t}[\hat{\rho}\left( 0\right)]$ cannot surpass that of the optimal initial state in a closed system. To address this, we must delve into the high-order corrections introduced by NPC noise. Perform the following substitution in Eq.~(\ref{Eq00}): $\hat{\mathcal{H}}\rightarrow\hat{\mathcal{H}}(\theta)$ and $\hat{\Pi}\left(t\right)\rightarrow \hat{\Pi}\left( \theta ,t\right)$. In this case, $\hat{\Pi}\left( \theta,t\right)$ is an effective dissipative superoperator containing the estimated parameter, implying an additional parameter encoding process beyond coherent dynamics.
This is essentially equivalent to the incoherent manipulation of quantum state by the NPC noise environment.
	Then a conclusion can be drawn that the estimation precision obtained from an open quantum system with NPC noise may surpass the precision limit determined by the optimal initial state and optimal estimation time of its closed counterpart, expressed as
	\begin{equation}
	\label{Eq8}
	F_{\theta }\left[ \hat{\rho}_{\theta}\left(\tau \right) \right]
	>F_{\theta }\left[ e^{\mathcal{\hat{H}}\left( \theta \right) t_{\text{optl}}}%
	\hat{\rho}^{\text{optl}}\left( 0\right) \right],
	\end{equation}%
	may hold. 
	Here, $\tau$ is a moment that depends on the specific form of the Hamiltonian and NPC noise. $\hat{\rho}^{\text{optl}}\left( 0\right) $ and $t_{\text{optl}}$ represent the optimal initial state and the optimal encoding time in the closed system, respectively.
	This surprising result contradicts intuition, as noise without estimated parameters can assist sensors in surpassing the precision limit established by coherent dynamics.
	Physically, this stems from NPC noise introducing an additional parameter encoding process, capitalizing on the non-commutativity between coherent and dissipative dynamics, compared to the noiseless case.

	\emph{Example}---Consider an open spin-$1/2$ system, whose dynamics is governed by a generalized master equation,
	\begin{equation}
	\label{Eq9}
	\frac{d\hat{\rho}\left( t\right) }{dt} =-i[\hat{H}_{\text{S}},\hat{\rho}\left( t\right) ]+\gamma[\hat{%
		\Gamma}\hat{\rho}\left( t\right) \hat{\Gamma}^{\dag } 
	-\frac{1}{2}\{\hat{\Gamma}^{\dag }\hat{\Gamma},\hat{\rho}\left( t\right) \}],
	\end{equation}
	where $\hat{H}_{\text{S}}=B\left[ \cos (\vartheta )\hat{\sigma}%
	_{x}+\sin (\vartheta )\hat{\sigma}_{z}\right] $ is the Hamiltonian of the system, with the Bohr magneton set to $1$ for simplicity. $B$ and $\vartheta $ denote the amplitude and direction of the magnetic field in the $ XZ$ plane, respectively. $\hat{\Gamma}=\cos (\alpha )\hat{\sigma}_{x}+\sin (\alpha )\hat{\sigma}_{z}$ is general quantum jump operator \cite{sun2016variational, WUYXSN}, where $\alpha$ is the coupling angle between the spin and environment bath. $\hat{\sigma}_{x,z}$ and $\gamma $ are the Pauli operator and the decay rate, respectively. 
	The model can be experimentally implemented using atoms or quantum dots \cite{NPCSSY,NPSCAX}, where atoms or electron spins are simultaneously excited and relaxed, accompanied by phase diffusion due to random fluctuations in the electromagnetic environment.
	We present three scenarios below. Scenario 1 highlights that PC noise degrades or has no impact on estimation precision. Scenario 2 demonstrates that NPC noise can enhance parameter estimation. Finally, Scenario 3 shows this enhancement has the potential to exceed the highest precision achievable in a noise-free environment.
	
	Scenario 1: when $\vartheta =\alpha =\pi /2$, the Hamiltonian $\hat{H}_{\text{S}}=B\hat{\sigma}_{z}$, the jump operator $\hat{\Gamma}=\hat{\sigma}_{z}$, and $[\hat H_{\text{S}}, \hat \Gamma]=0$ i.e., the system is affected by PC noise. 
	Suppose initial state of system is $\left\vert \Phi \left( 0\right) \right\rangle =\cos (\frac{%
		\beta }{2})\left\vert e\right\rangle +\sin (\frac{\beta }{2})\left\vert
	g\right\rangle $, where $\hat{\sigma}_{z}\left\vert e\right\rangle =\left\vert
	e\right\rangle $ and $\hat{\sigma}_{z}\left\vert g\right\rangle =-\left\vert
	g\right\rangle $. Let $\beta$ and $B$ be the estimated parameters, the QFI for each parameter reads \cite{XZTDD}
	\begin{subequations}
		\label{Eq11}
		\begin{eqnarray}
		F_{\beta }\left[ \hat{\rho}\left( t\right) \right] &=&1, 
		\\
		F_{B}\left[ \hat{\rho}\left( t\right) \right]  &=&4\sin ^{2}(\beta
		)e^{-4\gamma t}t^{2}. 
		\end{eqnarray}%
	\end{subequations}
	One can see that the PC noise has no effect on the estimation precision of $\beta $ but reduces that of $B$. These results are consistent with the conclusions presented earlier.

	Scenario 2: when $\vartheta =\pi /2$ and $\alpha \neq k\pi /2$ $(k\in \mathbb{Z})$,  $\hat{\Gamma}$ in this case signifies NPC noise. Suppose initial state is $\left\vert \Phi \left( 0\right) \right\rangle =\left\vert
	g\right\rangle $, the corresponding Bloch vector is $\vec{r}\left( 0\right) =%
	\left[ 0,0,-1\right] ^{\text{T}}$, lying on the negative $Z$ axis. 
	In the noiseless case, one can't get information of $B$ from evolved states because $\left\vert \Phi \left( t\right) \right\rangle =e^{iBt}\left\vert
	g\right\rangle $ and $F_{B}\left[\hat{\rho} (t)\right] =0$. 
	The non-commutative nature of NPC noise presents challenges in obtaining analytical expressions for $\hat\rho(t)$ and subsequently QFI. However, at the short-term limit,
	the Bloch vector of the system can be approximately solved~\cite{HZWK}, i.e, 
	\begin{equation}
	\vec{r}\left( t\right) =\Upsilon_\text{npc} \left( t\right) \vec{r}\left( 0\right) =-%
	\left[ \Upsilon _{13},\Upsilon _{23},\Upsilon _{33}\right] ^{\text{T}},
	\end{equation}%
	where $\Upsilon_{\text{npc}}$ represents the affine transformation matrix of the Bloch sphere, it signifies unequal contractions along the $X$, $Y$, and $Z$ axes, along with rotations around certain axes. Consequently, $\vec{r}(t)$ diverges from the $Z$ axis, acquiring quantum coherence and encoding effective information about $B$.
	The matrix elements $\Upsilon _{13}$ and $\Upsilon _{23}$ contain parameter $B$, leading
	to $F_{B}\left[ \hat{\rho}\left( t\right) \right] \neq 0$. This indicates that the NPC noise can boost quantum metrology.
	Furthermore, the presence of non-zero $\Upsilon _{13}$ and $\Upsilon _{23}$ indicates that the NPC noise imparts quantum coherence. This arises from the fact that the correlation between the decay channels through $\hat{\sigma}_{x}$ and $\hat{\sigma}_{z}$ in $\hat{\Gamma}$ is established by the dissipation process $\hat{\Gamma}\hat{\rho}(t)\hat{\Gamma}^\dagger$. 
	For PC noise, the affine transformation matrix $\Upsilon _{\text{pc}}$ shrinks the Bloch sphere equally along the $X$ and $Y$ axes and rotates around the $Z$ axis, making the Bloch vector $\vec{r}\left(t\right)$ always follow the $Z$ axis without containing effective information about $B$. See  Supplemental Material for specific forms of $\Upsilon _{\text{pc}}$  and $\Upsilon _{\text{npc}}$~\cite{HZWK}.
	Notice that, not all NPC noises can lead to the above results, e.g., $\hat{\Gamma}=\hat{\sigma}_{x}$.

	\begin{figure}[hbt!]
		\centering
		\includegraphics[width=0.94\linewidth]{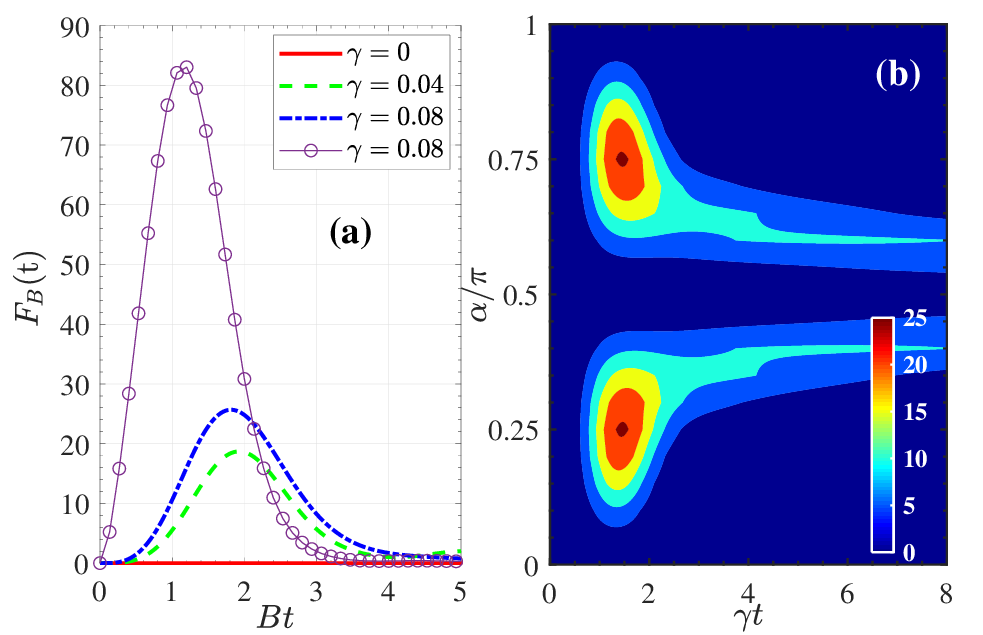}
		\caption{(a) $F_{B}$ versus encoding time $t$ with various decay rates, where $\alpha =\pi /4$ and $B=0.1$ (used as a scale).
			The purple circle line corresponds to the initial state $\left\vert \Phi \left( 0\right) \right\rangle =(\left\vert g\right\rangle+\left\vert e\right\rangle )/\sqrt{2}$ , while the others correspond to $\left\vert \Phi \left( 0\right) \right\rangle =\left\vert g\right\rangle $.
			(b) The density plot of $F_{B}$ versus $\alpha $ and $\gamma t$. }
		\label{fig1}
	\end{figure}
	
	We present the variation of $F_B$ based on the exact numerical solution of the master equation in Fig.~\ref{fig1}.
	Figure~\ref{fig1}(a) illustrates that NPC noise significantly enhances $F_B$, indicating enhanced estimation precision when the initial state is $\left\vert \Phi \left( 0\right) \right\rangle =\left\vert g\right\rangle $, and higher decay rates result in increased maximum value of $F_B$. But due to dissipation, $F_B$ eventually becomes zero over time.
	Interestingly, we observe that as the decay rate $\gamma$ increases, the value of $F_B$ derived from the initial state $\left\vert \Phi \left( 0\right) \right\rangle =\left\vert g\right\rangle $ temporarily surpasses the value achieved with the noise-free optimal state  $\left\vert \Phi \left( 0\right) \right\rangle =(\left\vert g\right\rangle+\left\vert e\right\rangle )/\sqrt{2}$ over a specific duration.
	This suggests a potential metrological advantage of non-optimal states in practical noisy environments.
	From Fig.~\ref{fig1}(b), we can see that the optimal coupling angle $\alpha_{\text{optl}}$ for the NPC noise-enhanced sensing precision is $\pi/4$ or $3\pi/4$. This is because when $\alpha =k\pi /4$ ($k$ is an odd number), the weights of $\hat \sigma_x$ and $\hat \sigma_z$ in the jump operator $\hat \Gamma$ are the same, maximizing the correlation between the two dissipation channels~\cite{HZWK}. In addition, Fig.~\ref{fig1}(b) exhibits symmetry with respect to $\alpha=\pi/2$, stemming from the fact that substituting $\alpha$ with $\pi-\alpha$ leaves the master Eq.~(\ref{Eq9}) unaffected.

	Scenario 3: Now, we consider the angle $\vartheta$ representing the direction of the magnetic field as the parameter to be estimated.
	We rewrite the Hamiltonian of system to $\hat{H}_{\text{S}}=\vec{R}\cdot \vec{J}$, where $\vec{R}=\left[ 2B\cos\left(\vartheta\right),0,2B\sin\left(\vartheta\right)\right]$ and $\vec{J}=\left[ \hat{\sigma}_{x}/2,0,\hat{\sigma}_{z}/2\right]$.
	Utilizing the method developed by Wang et al. \cite{PRAJXX}, the corresponding maximum QFI is given by
	\begin{equation}
	\label{S44}
	F_{\vartheta}^{\text{max}}\left(t\right) = \left( \frac{d|\vec{R}|}{%
		d\vartheta }\vec{e}_{R}\right) ^{2}t^{2}+4\left( \frac{d\vec{e}_{R}}{%
		d\vartheta }\right) ^{2}\text{sin}^{2}\left( \frac{|\vec{R}|}{2}t\right),
	\end{equation}
	where $|\vec{R}|$ and $\vec{e}_{R}$ are the magnitude and unit vector of $\vec{R}$, respectively.
	
	Since the magnitude of $\vec{R}$ is independent of $\vartheta$, the maximum noiseless QFI regarding $\vartheta$ expressed as
	$F_{\vartheta} =4\sin ^{2}\left( Bt\right)$
	which reaches the ultimate value of $4$ at the optimal encoding time $t_{\text{optl}}=\left(\pi/2+k\pi \right) /B$. 
	
	\begin{figure}[hbt!]
		\centering
		\includegraphics[width=1.02\linewidth]{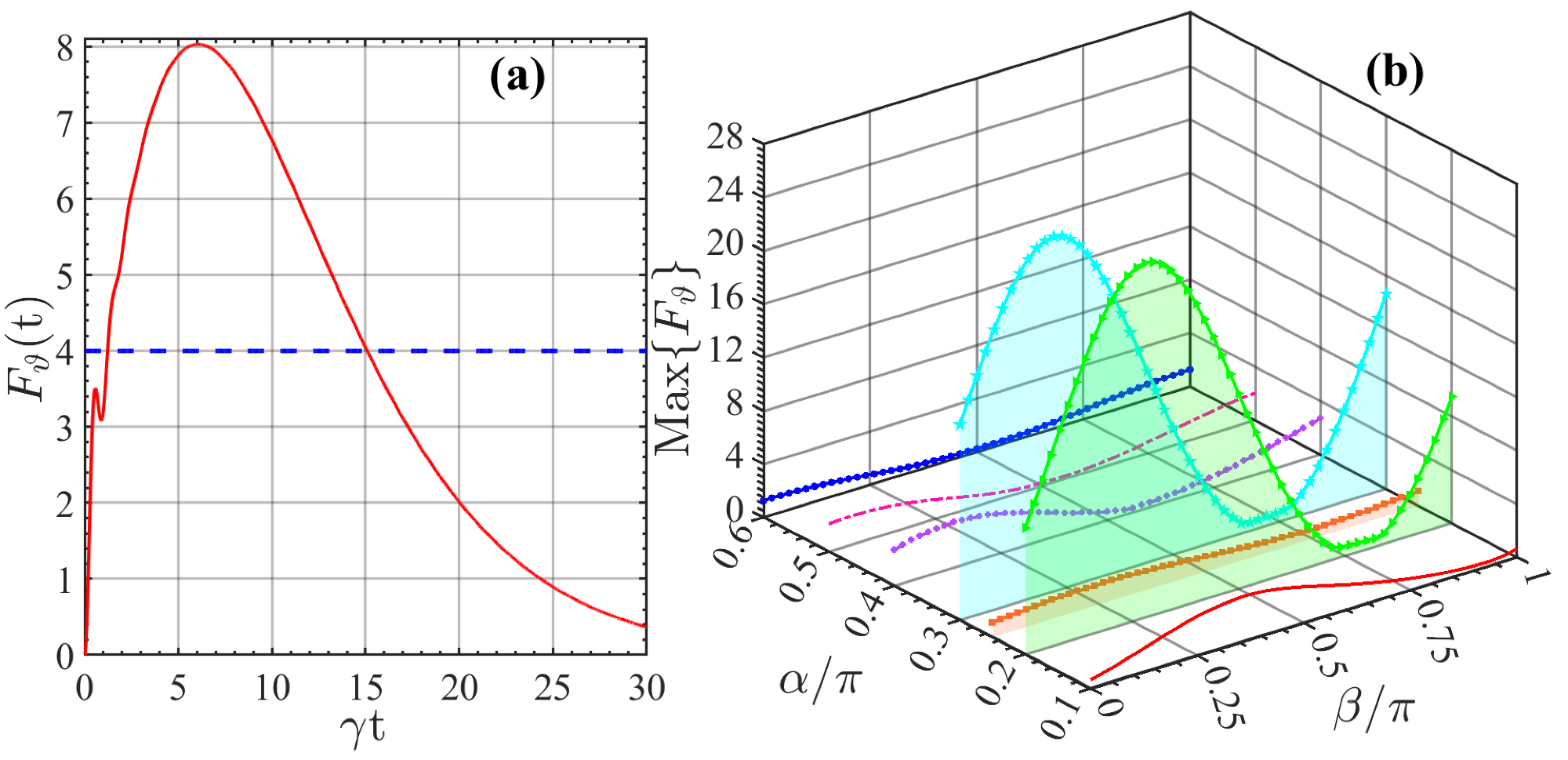}
		\caption{(a) $F_{\vartheta}$ (red solid line) versus encoding time $t$ in the presence of noise, where the magnetic field direction $\vartheta =\pi/3$, the noise coupling angle $\alpha=\pi /4$, $\beta =\pi /3$, $\gamma =0.03$ and  $B=0.1$ (used as a scale). The blue dashed line denotes the noise-free maximum of $F_\vartheta$. (b) The maximum QFI Max$\left\{ F_{\vartheta }\right\} $ as a function of $\alpha $ and $\beta $, where the magnetic field direction $\vartheta =\pi/4$,  the noise-free maximum QFI equals $4$, and other parameters are the same as (a). The orange curve corresponds to $\alpha=\vartheta$. }
		\label{fig2}
	\end{figure}
	
	The primary effect of NPC noise on the QFI can also be demonstrated within this model by utilizing the reaction-coordinate polaron transform to introduce an effective Hamiltonian $\hat{H}_{\text{S}}^{\text{eff}}=\vec{R}^\prime \cdot \vec{J}$~\cite{zjl}. This Hamiltonian accurately captures the dominant dynamics of the system within the open environment.
	We observe that unlike $\vec{R}$, both the magnitude and direction of $\vec{R}'$ vary with $\vartheta$. Especially, the change in its magnitude leads to an accelerated increase in $F_\vartheta^\text{max}$ with a factor of $t^2$, far surpassing the $\sin^2(B t)$ factor derived from directional changes, indicating a potential to exceed a maximum of noiseless QFI.
	
	Fig.~\ref{fig2}(a) shows the numerical simulation of $F_{\vartheta}$  evolving over time with noise based on the exact quantum master equation, exceeding 4 for a specific duration. This verifies our theory that NPC noise can enhance the non-Hermitian sensor's precision in measuring magnetic field direction beyond its Hermitian counterpart's limit.
	Fig.~\ref{fig2}(b) presents a 3D plot depicting the maximum QFI Max$\left\{ F_{\vartheta }\right\} $ as a function of both noise coupling angle $\alpha$ and initial state parameter $\beta$ in the noisy environment. These maximums represent the peaks throughout time evolution under given initial states, rather than the maximum QFI $F_{\vartheta}^{\text{max}}\left(t\right)$  given in the optimal initial state. 
	The plot suggests that if there is a substantial difference between $\alpha$ and $\vartheta$, surpassing the precision limit is impossible, regardless of the chosen initial state.
	In contrast, when $\alpha$ is very close to $\vartheta$, it becomes more feasible to surpass the precision limit by selecting an appropriate initial state. 
	However, this comes at the expense of requiring a longer encoding time. Fortunately, the open system takes a considerable time to decay to a steady state in this case, thereby affording an extended window for encoding \cite{wsl}. But for the special case of $\alpha=\vartheta$, indicating the transition from NPC to PC noise, the highest precision limit set by coherent dynamics cannot be exceeded.

	\begin{figure}[hbt!]
		\centering
		\includegraphics[width=0.9\linewidth]{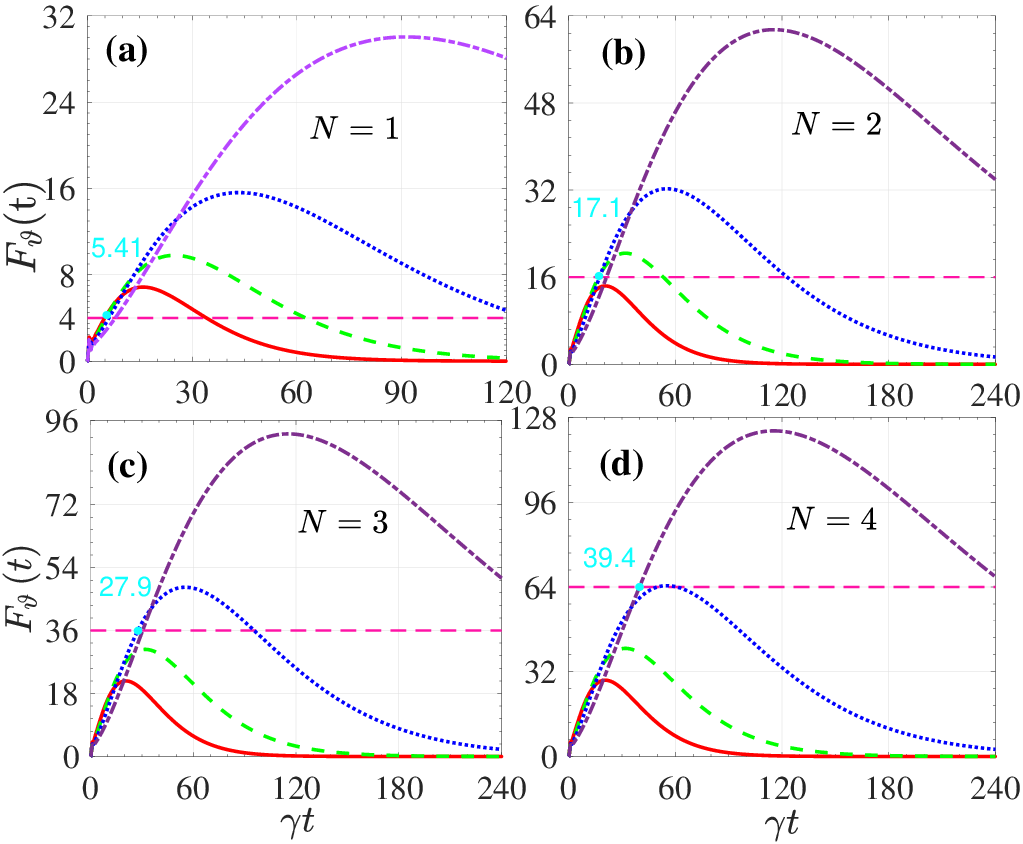}
		\caption{$F_{\vartheta}$ versus $\gamma t$ for different coupling angles $\alpha$, as the particle number $N$ increases, with magnetic field direction $\vartheta=\pi/3$. The red solid, green dashed, blue dotted, and purple dotted-dashed lines correspond to $\alpha=0.28\pi, 0.29\pi, 0.3\pi, 0.31\pi$, respectively. The pink dashed lines in each subplot represent the maximum QFI set by the coherent dynamics, while the sky blue numbers indicate the time it takes to surpass them.}
		\label{fig3}
	\end{figure} 
	
	\emph{Multi-particle scenario}---The results obtained in the previous section should also hold for collective systems composed of $N$ particles, as Eq.~(\ref{Eq00}) is universal and does not confine the analysis to a specific model. To verify this, we simulated the corresponding $N$-particle master equation and computed the QFI~\cite{WLKS}, assuming no direct coupling between particles for generality.
	Fig.~\ref{fig3} plots the variation of $F_{\vartheta}$ with $\gamma t$ for different coupling angles $\alpha$, where the initial state of the system is $\left\vert \Phi _{\text{tot}}\left( 0\right) \right\rangle =\left[ \left( \left\vert e\right\rangle +\left\vert g\right\rangle \right)/\sqrt{2}
	\right] ^{\otimes N}$ (For entangled initial states, see  Supplemental Material for similar results.). We observe that for particle number $N=2, 3, 4$, as long as the coupling angle $\alpha$ is chosen appropriately, and with the aid of NPC noise, the estimation precision consistently exceeds the ultimate precision limit set by the optimal initial state (entangled state) in the absence of noise. However, it is worth noting that as $N$ increases, surpassing this limit through the introduction of NPC noise becomes progressively more challenging.
	This is because the selection of the coupling angle $\alpha$ becomes more stringent (manifested as $\alpha$  needing to be closer to $\vartheta$), and the encoding time also becomes longer (see the sky blue numbers), posing significant challenges for practical experimental realization. Notice that, in the absence of NPC noise, the maximum value of $F_\vartheta$ in an $N$-particle system is $4 N^2$ (see the pink dashed lines). This implies that the precision limit can reach the Heisenberg scale in terms of particle number. Although NPC noise cannot break this scale, it can, in principle, make the $F_\vartheta$ value exceed $4 N^2$.
	
	\emph{Discussion and summary}---More recently, a study reported that  non-Hermitian sensors do not
	outperform their Hermitian counterparts
	in the performance of sensitivity~\cite{CSWXG}. The authors derived an upper
	bound of channel QFI, i.e.,
	$F_{\lambda }^{\left( c\right) }\left( t\right) \leq \left[
	\int\nolimits_{0}^{t}|| \partial \hat{H}_{\lambda }\left( s\right)
	/\partial \lambda || ds\right] ^{2}=  \mathcal{F}_{\lambda}^{\text{UB}}$, where $\hat{H}_{\lambda
	}\left( t\right) $ is the parameter-dependent Hamiltonian of the system, and $F_{\lambda }^{\left( c\right)
	}\left( t\right) $ is the maximum QFI achievable by optimizing the initial state.  We point out that the equality sign in the above inequality can
	be achieved when the estimated parameter is the overall factor of the  Hamiltonian \cite{hsds}, e.g., $\hat{H}_{B } =B \hat{%
		\sigma}_{z}\Rightarrow F_{B }^{\left( c\right) }\left( t\right)=\mathcal{F}_{B}^{\text{UB}}=4t^{2}$.  
	However, the estimated parameter is not always an overall multiplicative factor of the Hamiltonian, e.g., $\hat{H}_{\vartheta } = B\left[\cos( \vartheta) \hat{\sigma}_{x}+\sin( \vartheta) \hat{\sigma}_{z}\right]$. In this case, $F_{\vartheta }^{\left( c\right) }\left( t\right) =4\sin^{2}\left(Bt\right)$, well below the up bound $\mathcal{F}_{\vartheta}^{\text{UB}}=4B^2t^2$ for $t\gg 1$. Our study suggests that appropriate NPC noise can bridge this gap and enhance the precision limit, without violating the inequality.

	In summary, we found that NPC noise can enhance quantum metrology due to the non-commutative nature between coherent and dissipative dynamics. Remarkably, the QFI attained through dynamics influenced by NPC noise can surpass the ultimate limit set solely by coherent dynamics. This suggests that the sensing precision of a non-Hermitian sensor with NPC noise can potentially outperform its Hermitian counterpart. 
	Utilizing a general series solution analysis of the master equation, we establish the universality of our findings and demonstrate a specific instance of noise-enhanced magnetic-field quantum metrology. Furthermore, this approach provides valuable insights for enhancing other quantum technologies that are constrained by environmental noise.

	\begin{acknowledgments}
		\emph{Acknowledgements}---This work was supported by the
		Innovation Program for Quantum Science and
		Technology (No. 2021ZD0303200); the National Science
		Foundation of China (No. 12374328, No. 11974116, and
		No. 12234014); the Shanghai Municipal Science and
		Technology Major Project (No. 2019SHZDZX01); the
		National Key Research and Development Program of
		China (No. 2016YFA0302001); the Fundamental
		Research Funds for the Central Universities; the Chinese
		National Youth Talent Support Program, and the Shanghai
		Talent program.
	\end{acknowledgments}
	
	\nocite{*}
	\bibliography{Ref}
\appendix
\begin{widetext}
\section{Enhanced Quantum Metrology with Non-Phase-Covariant Noise: \\ Supplemental Material} \label{A}
\section{I. Series solution of the master equation}
A powerful tool to describe the dynamics of open systems is the convolutionless master equation,
i.e., \cite{rivas2012open,breuer2002theory} 
\begin{equation}
\label{es1}
\partial _{t}\hat{\rho}\left( t\right) =( \hat{\mathcal{H}}+\xi %
\hat{\mathcal{L}}) \hat{\rho}\left( t\right) ,  
\end{equation}%
with 
\begin{equation}
\label{es2}
\hat{\mathcal{H}} =-i[ \hat{H},\bullet ] , \hat{\mathcal{L}} =\sum_{k}\gamma _{k}[ \hat{\Gamma}_{k}\bullet \hat{%
	\Gamma}_{k}^{\dag }-\frac{1}{2}\{ \hat{\Gamma}_{k}^{\dag }\hat{\Gamma}%
_{k},\bullet \} ],  
\end{equation}%
where $\hat{\rho}\left( t\right) $ denotes the reduced density matrix of
the system, $\xi $ is a perturbation parameter which can be roughly
considered as describing the coupling strength between the system and the
environment.
The superoperators $\hat{\mathcal{H}}$ and $\hat{\mathcal{L}}$ act on the transformed operator (denoted by the symbol “$\bullet $”) according to Eq.~(\ref{es2}), where $\hat{H}$ represents the system's Hamiltonian and $\hat{\Gamma}_{k}$ is the quantum jump operator associated with a damping channel characterized by the decay rate $\gamma _{k}$.
Based on
Eq.~(\ref{es1}), it is easy to get the formal solution as 
\begin{equation}
\hat{\rho}\left( t\right) =e^{\left( \hat{\mathcal{H}}+\xi \mathcal{\hat{%
			L}}\right) t}\hat{\rho}\left( 0\right) =\sum_{n=0}^{\infty }\frac{%
	t^{n}\left( \hat{\mathcal{H}}+\xi \mathcal{\hat{L}}\right) ^{n}}{n!}\hat{\rho%
}\left( 0\right) ,  \label{es3}
\end{equation}
where $\hat{\rho}\left( 0\right) $ represents the initial state of the system. 

If the coupling between the system and the environment is weak, namely $\xi \gamma _{k}$ is much smaller than the characteristic frequency of the system, then focusing on the first-order correction of the environment to the system evolution is sufficient. By applying the binomial theorem to Eq.~(\ref{es3}) and retaining the first-order terms in $\xi$, one can obtain
\begin{equation}
\hat{\rho}\left( t\right) \approx \left[ \sum_{n=0}^{\infty }\frac{t^{n}%
	\hat{H}^{n}}{n!}+\xi \sum_{n=1}^{\infty }\frac{t^{n}}{n!}\sum_{m=0}^{n-1}%
\hat{\mathcal{H}}^{m}\mathcal{\hat{L}}\hat{\mathcal{H}}^{n-m-1}\right] \hat{%
	\rho}\left( 0\right) .  \label{es4}
\end{equation}%
To simplify Eq.~(\ref{es4}), we define an $2\times 2$ array $\mathbf{\hat{\Pi}}%
_{2}$ composed of superoperators, its specific form reads as \cite{villegas2016application}
\begin{equation}
\label{es5}
\mathbf{\hat{\Pi}}_{2}=\left[ 
\begin{array}{cc}
\hat{\mathcal{H}} & \mathcal{\hat{L}} \\ 
0 & \hat{\mathcal{H}}%
\end{array}%
\right] .
\end{equation}%
Through straightforward algebraic manipulations, one can confirm the validity of the following relationship
\begin{equation}
\label{es6}
\left( \mathbf{\hat{\Pi}}_{2}^{n}\right) _{12}=\sum_{m=0}^{n-1}\hat{\mathcal{%
		H}}^{m}\mathcal{\hat{L}}\hat{\mathcal{H}}^{n-m-1},
\end{equation}%
where the subscript “$12$” denotes the element in the first row and the second column of the array. Therefore, Eq.~(\ref{es4}) can be rewritten as 
\begin{equation}
\hat{\rho}\left( t\right) \approx  \left[ e^{\mathcal{\hat{H}}t}+\xi \left( e^{\mathbf{\hat{\Pi}}%
	_{2}t}\right) _{12}\right] \hat{\rho}\left( 0\right) . \label{es7}
\end{equation}%
Alternatively, one can divide $\hat{\rho}\left( t\right) $ into two terms in the form 
\begin{equation}
\ \hat{\rho}\left( t\right) \approx \hat{\rho}^{\left( 0\right)
}\left( t\right) +\xi \left[ \hat{\rho}_{A}^{\left( 1\right) }\left(
t\right) \right] _{12},  \label{es8}
\end{equation}%
where the first term represents the system's evolution under the Hamiltonian $\hat{H}$, while the second term denotes the first-order correction due to noise on the system's evolution, with 
\begin{equation}
\label{es9}
\hat{\rho}_{A}^{\left( 1\right) }\left( t\right) =\left[ 
\begin{array}{cc}
\hat{\rho}_{A,11}^{\left( 1\right) }\left( t\right) & \hat{\rho}%
_{A,12}^{\left( 1\right) }\left( t\right) \\ 
\hat{\rho}_{A,21}^{\left( 1\right) }\left( t\right) & \hat{\rho}%
_{A,22}^{\left( 1\right) }\left( t\right)%
\end{array}%
\right] .
\end{equation}%
Note that here and later, the superscript ``($i$)''
indicates the modified order. 
By taking the derivative of Eqs.~(\ref{es7})
and (\ref{es8}) with respect to time and comparing the coefficients of each power of $\xi$, one can see that
\begin{equation}
\label{es10}
\frac{d\hat{\rho}_{A}^{\left( 1\right) }\left( t\right) }{dt}=\mathbf{\hat{%
		\Pi}}_{2}\hat{\rho}_{A}^{\left( 1\right) }\left( t\right), 
\end{equation}%
with the initial condition
\begin{equation}
\label{es11}
\hat{\rho}_{A}^{\left( 1\right) }\left( 0\right) =\left[ 
\begin{array}{cc}
\hat{\rho}\left( 0\right) & 0 \\ 
0 & \hat{\rho}\left( 0\right)%
\end{array}%
\right] .
\end{equation}%
Based on the form of $\mathbf{\hat{\Pi}}_{2}$, one can further obtain the following differential equations
\begin{subequations}
	\label{es12}
	\begin{eqnarray}
	\frac{d\hat{\rho}_{A,12}^{\left( 1\right) }\left( t\right) }{dt} &=&\hat{%
		\mathcal{H}}\hat{\rho}_{A,12}^{\left( 1\right) }\left( t\right) +\mathcal{%
		\hat{L}}\hat{\rho}_{A,22}^{\left( 1\right) }\left( t\right) , \\
	\frac{d\hat{\rho}_{A,22}^{\left( 1\right) }\left( t\right) }{dt} &=&\hat{%
		\mathcal{H}}\hat{\rho}_{A,22}^{\left( 1\right) }\left( t\right) ,
	\end{eqnarray}%
\end{subequations}
where $\hat{\rho}_{A,12}^{\left( 1\right) }\left( 0\right) =0$ and $\hat{\rho%
}_{A,22}^{\left( 1\right) }\left( 0\right) =\hat{\rho}\left( 0\right).$
By solving differential Eqs.~(\ref{es12}) , one can obtain  
\begin{equation}
\label{es13}
\hat{\rho}_{A,12}^{\left( 1\right) }\left( t\right) =e^{\mathcal{\hat{H}}t}%
\left[ \int_{0}^{t}e^{-\mathcal{\hat{H}}t_{1}}\mathcal{\hat{L}}e^{\mathcal{%
		\hat{H}}t_{1}}dt_{1}\right] \hat{\rho}\left( 0\right) .
\end{equation}%
Therefore, considering the first-order correction of noise to the unperturbed solution, the reduced density matrix of the system reads
\begin{equation}
\label{es14}
\hat{\rho}\left( t\right) \approx e^{\mathcal{\hat{H}}t}\left[ \hat{\mathbb{1}}+\xi
\int_{0}^{t}e^{-\mathcal{\hat{H}}t_{1}}\mathcal{\hat{L}}e^{\mathcal{\hat{H}}%
	t_{1}}dt_{1}\right] \hat{\rho}\left( 0\right).
\end{equation}%

Based on the same way, we can derive the exact solution of the master equation containing higher-order corrections by constructing superoperator array,
\begin{equation}
\label{es15}
\mathbf{\hat{\Pi}}_{n}=\left[ 
\begin{array}{ccccc}
\mathcal{\hat{H}} & \mathcal{\hat{L}} & \cdots & 0 & 0 \\ 
0 & \mathcal{\hat{H}} & \mathcal{\hat{L}} & \cdots & 0 \\ 
\vdots & 0 & \ddots & \ddots & \vdots \\ 
\vdots & 0 & 0 & \ddots & \mathcal{\hat{L}} \\ 
0 & \vdots & 0 & 0 & \mathcal{\hat{H}}%
\end{array}%
\right] .
\end{equation}%
Then the $n$th-order correction due to noise on the system's density matrix is 
\begin{equation}
\label{es16}
\xi ^{n}\left[ \hat{\rho}_{A}^{\left( n\right) }\left( t\right) \right]
_{1n+1}=\xi ^{n}e^{\mathcal{\hat{H}}t}\left[\int_{0}^{t}\int_{0}^{t_{1}}\cdots
\int_{0}^{t_{n-1}}e^{-\mathcal{\hat{H}}t_{1}}\mathcal{\hat{L}}e^{\mathcal{%
		\hat{H}}t_{1}}e^{-\mathcal{\hat{H}}t_{2}}\mathcal{\hat{L}}e^{\mathcal{\hat{H}%
	}t_{2}}\cdots e^{-\mathcal{\hat{H}}t_{n}}\mathcal{\hat{L}}e^{\mathcal{\hat{H}%
	}t_{n}}d{t_{n}}d{t_{n-1}}\ldots d{t_{2}}d{t_{1}}\right]\hat{\rho}\left(
0\right),
\end{equation}%
where  the integral upper limit satisfied $t>t_{1}>\cdots >t_{n-2}>t_{n-1}$ in the
time-ordering integral. 
Then the exact form of the system's density matrix can be expressed in the following series form 
\begin{equation}
\label{es17}
\hat{\rho}\left( t\right) =\sum_{n=0}^{\infty }\xi ^{n}\left[ \hat{\rho}%
_{A}^{\left( n\right) }\left( t\right) \right] _{1n+1},
\end{equation}%
in which $\left[ \hat{\rho}_{A}^{\left( 0\right) }\left( t\right) \right]
_{11}=\hat{\rho}^{\left( 0\right) }\left( t\right).$ Particularly, in
the case of $\xi =1$, one can get   
\begin{eqnarray}
\label{es18}
\hat{\rho}\left( t\right) &=&\sum_{n=0}^{\infty }\left[ \hat{\rho}%
_{A}^{\left( n\right) }\left( t\right) \right] _{1n+1}  \notag \\
&=&e^{\mathcal{\hat{H}}t}\left[ \hat{\mathbb{1}}+\int_{0}^{t}e^{-\mathcal{\hat{H}}t_{1}}%
\mathcal{\hat{L}}e^{\mathcal{\hat{H}}t_{1}}dt_{1}+\int_{0}^{t}%
\int_{0}^{t_{1}}e^{-\mathcal{\hat{H}}t_{1}}\mathcal{\hat{L}}e^{\mathcal{\hat{%
			H}}t_{1}}e^{-\mathcal{\hat{H}}t_{2}}\mathcal{\hat{L}}e^{\mathcal{\hat{H}}%
	t_{2}}d_{t_{2}}d_{t_{1}}+\ldots \right] \hat{\rho}\left( 0\right),
\end{eqnarray}
which is Eq.~(1) in the main text.

For PC-noise, the dynamics induced by the noise terms commute with that caused by the system Hamiltonian, i.e., $\mathcal{\hat{H}}[ \mathcal{\hat{L}%
}[\hat{\rho}\left( t\right)]] =\mathcal{\hat{L}}[ \mathcal{\hat{H}}[\hat{\rho}\left( t\right)]] $ and $e^{-\mathcal{\hat{H}}t_{k}}\mathcal{\hat{L}}e^{\mathcal{\hat{H}}t_{k}}=\mathcal{\hat{L}}$, so Eq.~(\ref{es18}) can be reduced to
\begin{eqnarray}
\hat{\rho}\left( t\right) &=&e^{\mathcal{\hat{H}}t}\left[ \hat{\mathbb{1}}+\mathcal{\hat{L}}%
t+\cdots +\frac{t^{n}\mathcal{\hat{L}}^{n}}{n!}+\cdots \right] \hat{\rho}\left( 0\right)  \notag \\
&=&e^{\mathcal{\hat{H}} t}[e^{\mathcal{\hat{L}}t}[\hat{%
	\rho}\left( 0\right)]]  \notag \\
&=&e^{\mathcal{\hat{L}}t}[\hat{\rho}^{\left( 0\right) }\left( t\right)], \label{es19} 
\end{eqnarray}%
which is Eq.~(3) in the main text. Here $\hat{\rho}%
^{\left( 0\right) }\left( t\right) $ denotes the state obtained by $
\mathcal{\hat{H}}$ parameterizing the initial state.

In the case of NPC-noise, the equation $\mathcal{\hat{H}}[ \mathcal{\hat{L}}[\hat{\rho}\left( t\right)]] =\mathcal{\hat{L}}[ \mathcal{\hat{H}}[\hat{\rho}\left( t\right)]] $ no longer holds. Moreover, since $\mathcal{\hat{H}}$ and $\mathcal{\hat{L}}$ are superoperators, obtaining a concise and precise expression for $\hat{\rho}(t)$ as in the previous case is challenging. However, we can derive a similar approximated expression when the evolution time is short. 
In the case of short time, the expansion of $e^{-\mathcal{\hat{H}}t_{k}}\mathcal{\hat{L}}e^{\mathcal{\hat{H}}t_{k}}$ can be approximately written as~\cite{griffiths2018introduction} 
\begin{equation}
e^{-\mathcal{\hat{H}}t_{k}}\mathcal{\hat{L}}e^{\mathcal{\hat{H}}t_{k}}=%
\mathcal{\hat{L}}-t_{k}[ \mathcal{\hat{H}},\mathcal{\hat{L}}] +%
\frac{t_{k}^{2}}{2!}[ \mathcal{\hat{H}},[ \mathcal{\hat{H}},%
\mathcal{\hat{L}}]] +\cdots \approx \mathcal{\hat{L}+}o\left( t_{k}\right) . \label{es23}
\end{equation}%
Further, the following approximate relationships are established 
\begin{subequations}
	\begin{align}
	e^{-\mathcal{\hat{H}}t_{1}}\mathcal{\hat{L}}e^{\mathcal{\hat{H}}t_{1}}
	&\approx \mathcal{\hat{L}+}o\left( t_{1}\right) , \\
	e^{-\mathcal{\hat{H}}t_{1}}\mathcal{\hat{L}}e^{\mathcal{\hat{H}}t_{1}}e^{-%
		\mathcal{\hat{H}}t_{2}}\mathcal{\hat{L}}e^{\mathcal{\hat{H}}t_{2}} &\approx %
	\mathcal{\hat{L}}^{2}+o\left( t_{2}\right) , \\
	&\vdots  \notag \\
	e^{-\mathcal{\hat{H}}t_{1}}\mathcal{\hat{L}}e^{\mathcal{\hat{H}}t_{1}}e^{-%
		\mathcal{\hat{H}}t_{2}}\mathcal{\hat{L}}e^{\mathcal{\hat{H}}t_{2}}\cdots e^{-%
		\mathcal{\hat{H}}t_{m}}\mathcal{\hat{L}}e^{\mathcal{\hat{H}}t_{m}} &\approx %
	\mathcal{\hat{L}}^{m}+o\left( t_{m}\right) .
	\end{align}%
\end{subequations} 
Finally, Eq.~(\ref{es18}) can be approximated as 
\begin{align}
\hat{\rho}\left( t\right) 
&\approx e^{\mathcal{\hat{H}} t}\left[
\hat{\mathbb{1}}+\int_{0}^{t}\mathcal{\hat{L}}dt_{1}+\int_{0}^{t}\int_{0}^{t_{1}}%
\mathcal{\hat{L}}^{2}d_{t_{2}}d_{t_{1}}+\ldots \right] \hat{\rho}\left(
0\right)  \notag \\
&=e^{\mathcal{\hat{H}} t}[e^{\mathcal{\hat{L}}t}[\hat{%
	\rho}\left( 0\right)]], \label{es29}
\end{align}%
which is Eq.~(4) in the main text. It should be noted that, while the form of Eq.~(\ref{es29}) is similar to Eq.~(\ref{es19}), this form is only valid in the short-time regime, and $\mathcal{\hat H}$ and $\mathcal{\hat L}$ do not commute.

\section{II. Derivation of Eq.~(5) in the main text}
The dynamic evolution of a $d$-dimensional open quantum system is described
by the master equation
\begin{equation}
\label{EqS1}
\frac{\partial \hat{\rho}(t)}{\partial t}=(\hat{\mathcal{H}}+\hat{\mathcal{L}})[ \hat{\rho}%
(t)],
\end{equation}%
with 
\begin{equation}
\hat{\mathcal{H}}\left[ \bullet \right] =-i[\hat{H},\bullet ],\hat{\mathcal{L}}\left[ \bullet %
\right] =\sum_{k}\gamma _{k}[\Gamma _{k}\bullet \Gamma _{k}^{\dag }-\frac{1}{%
	2}\{\Gamma _{k}^{\dag }\Gamma _{k},\bullet \}].
\end{equation}

For noiseless situation, the Eq.~(\ref{EqS1}) is reduced to
\begin{equation}
\label{EqS3}
\frac{\partial \hat{\rho}^{\left( 0\right) }(t)}{\partial t}=\hat{\mathcal{H}}[ 
\hat{\rho}^{\left( 0\right) }(t)] =-i[\hat{H},\hat{\rho}^{\left(
	0\right) }(t)],
\end{equation}%
where $\hat{\rho}^{\left( 0\right) }(t)$ represents the evolved state in the
noiseless case, and the initial condition is $\hat{\rho}^{\left( 0\right)
}(0)=\hat{\rho}(0)$. To solve the Eq.~(\ref{EqS3}), we consider
eigenequations of superoperator $\hat{\mathcal{H}}$ \cite{MFACD,PRL104,PJX1}
\begin{subequations}
	\begin{eqnarray}
	\hat{\mathcal{H}}\hat{\chi}_{i}^{R} &=&E_{i}\hat{\chi}_{i}^{R}, \\
	\hat{\mathcal{H}}^{\dag }\hat{\chi}_{i}^{L} &=&E_{i}^{\ast }\hat{\chi}_{i}^{L},
	\end{eqnarray}%
\end{subequations}
where $E_{i}$ $(E_{i}^{\ast })$ and $\hat{\chi}_{i}^{R}$ $\left( \hat{\chi}%
_{i}^{L}\right) $ denote the eigenvalues and corresponding right (left)
eigenmatrices of $\hat{\mathcal{H}}$ ($i=1\sim d^{2}$). Here, we do not consider the
case that $\hat{\chi}_{i}^{R}$ $\left( \hat{\chi}_{i}^{L}\right) $ coalesce,
and such a situation is believed to be rare \cite{PRL104}. In particular, different left (right) eigenmatrices may be not orthogonal due to
superoperator $\hat{\mathcal{H}}$ being non-Hermitian, i.e., Tr$[ \hat{%
	\chi}_{i}^{R(L)^{\dag }}\hat{\chi}_{j}^{R(L)}] \neq \delta _{ij}$ ($%
\delta _{ij}$ is Kronecker delta function) \cite{MFACD,PRL104,PJX1},  while  the left
and right eigenmatrices with different eigenvalues may be orthogonal to each
other, namely Tr$[ \hat{\chi}%
_{i}^{L^{\dag }}\hat{\chi}_{j}^{R}] =\delta _{ij}$. Importantly, despite different right eigenmatrices $\hat{\chi}%
_{i}^{R}$ are not orthogonal to each other, they can still form a complete
basis $\{\hat{\chi}_{i}^{R}\}$. Thus, the density matrix $\hat{\rho}^{\left(
	0\right) }(t)$ can be written as \cite{MFACD,PRL104,PJX1}
\begin{equation}
\hat{\rho}^{\left( 0\right) }(t)=\sum_{i=1}^{d^{2}}\text{Tr}[ \hat{\chi}%
_{i}^{L^{\dag }}\hat{\rho}(0)] e^{E_{i}t}\hat{\chi}_{i}^{R},
\end{equation}%
where $\hat{\rho}^{\left( 0\right) }(0)=\hat{\rho}(0)$ has been utilized. 

Next, we introduce environment noise. For the Liouvillian superoperator $\hat{\mathcal{L}}$,
we can also construct its eigenequations \cite{MFACD,PRL104,PJX1}
\begin{subequations}
	\label{BZFDC}
	\begin{eqnarray}
	\hat{\mathcal{L}}\hat{\Re}_{j}^{R} &=&\zeta _{j}\hat{\Re}_{j}^{R}, \\
	\hat{\mathcal{L}}^{\dag }\hat{\Re}_{j}^{L} &=&\zeta _{j}^{\ast }\hat{\Re}_{j}^{L},
	\end{eqnarray}%
\end{subequations}
where $\zeta _{j}$ $(\zeta _{j}^{\ast })$ and $\hat{\Re}_{j}^{R}$ $(\hat{\Re}%
_{j}^{L})$ denote the eigenvalues and corresponding right (left)
eigenmatrices of $\hat{\mathcal{L}}$ ($j=1\sim d^{2}$). Notice that Liouvillian
superoperator is negative semidefinite when $\gamma _{k}\geq 0$ for $\forall
k$, i.e., Re$[\zeta _{j}]\leq 0$ \cite{breuer2002theory}. Similar to $\hat{\mathcal{H}}$, $\hat{\mathcal{L}}$'s
eigenmatrices  can also construct a complete basis  $\{\hat{\Re}_{j}^{R}\}$ \cite{MFACD,PRL104,PJX1}. In order to better compare the output states with and without noise, we  expand $\hat{\chi}_{i}^{R}$ through the eigenmatrices $\{\hat{\Re}_{j}^{R}\}$ of $\hat{\mathcal{L}}$, i.e., \cite{MFACD,PRL104,PJX1}
\begin{equation}
\hat{\chi}_{i}^{R}=\sum_{j=1}^{d^{2}}\text{Tr}[ \hat{\Re}_{j}^{L^{\dag }}%
\hat{\chi}_{i}^{R}] \hat{\Re}_{j}^{R}.
\end{equation}%
In this case, $\hat{\rho}^{\left( 0\right) }(t)$ can be rewritten as
\begin{equation}
\hat{\rho}^{\left( 0\right) }(t)=\sum_{i=1}^{d^{2}}\sum_{j=1}^{d^{2}}\text{Tr%
}[ \hat{\chi}_{i}^{L^{\dag }}\hat{\rho}(0)] \text{Tr}[ \hat{\Re%
}_{j}^{L^{\dag }}\hat{\chi}_{i}^{R}] e^{E_{i}t}\hat{\Re}_{j}^{R}.
\end{equation}%
For PC noise, we have that
\begin{equation}
\hat{\rho}(t) =e^{\hat{\mathcal{L}}t}[ \hat{\rho}^{\left( 0\right) }(t)] 
=\sum_{i=1}^{d^{2}}\sum_{j=1}^{d^{2}}\text{Tr}[ \hat{\chi}%
_{i}^{L^{\dag }}\hat{\rho}(0)] \text{Tr}[ \hat{\Re}_{j}^{L^{\dag }}%
\hat{\chi}_{i}^{R}] e^{ E_{i} t}\hat{\Re}%
_{j}^{R}e^{\zeta _{j} t},
\end{equation}%
where we have used Eqs.~(\ref{es19}) and (\ref{BZFDC}).
Notice that the density matrix  $\hat{\rho}(t)$
has an additional parameter-independent exponential decay factor $e^{\zeta _{j}t}$ compared with
the noise-free evolved state $\hat{\rho}^{\left( 0\right) }(t)$ due to Re$[\zeta _{j}]\leq 0$. Obviously,
this parameter-independent exponential decay factor $e^{\zeta _{j}t}$ will cause a potential
decrease in the elements of the density matrix  and information of parameters
to be estimated (this does not destroy $\text{Tr}[\hat{\rho}(t)]=1$)  \cite{rivas2012open,breuer2002theory} . Thus, for PC noise,  we can conclude that
\begin{equation}
F_{\theta }\left[ \hat{\rho}_{\theta }(t)\right] \leq F_{\theta }[\hat{\rho}%
_{\theta }^{\left( 0\right) }(t)].
\end{equation}%
This is Eq.~(5) in the main text.

\section{III. Derivation of Eq.~(8) in the main text}
The initial state of the system is $|\Phi \left( 0\right) \rangle =$ cos$%
\left( \tfrac{\beta }{2}\right) |e\rangle +$ sin$\left( \tfrac{\beta }{2}%
\right) |g\rangle $, and in the $\hat{\sigma}_{z}$ representation the corresponding density matrix is
\begin{equation}
\hat{\rho}\left( 0\right) =\left[ 
\begin{array}{cc}
\text{cos}^{2}\left( \tfrac{\beta }{2}\right)  & \frac{1}{2}\text{sin}\left(
\beta \right)  \\ 
\frac{1}{2}\text{sin}\left( \beta \right)  & \text{sin}^{2}\left( \tfrac{%
	\beta }{2}\right) 
\end{array}%
\right] .
\end{equation}%
The analytical solution for the
master equation,
\begin{equation}
\frac{d\hat{\rho}\left( t\right) }{dt}=-i\left[ B\hat{\sigma}_{z},\hat{\rho}\left(
t\right) \right] +\gamma \left[ \hat{\sigma}_{z}\hat{\rho}\left( t\right) 
\hat{\sigma}_{z}-\hat{\rho}\left( t\right) \right],
\end{equation}%
is
\begin{equation}
\hat{\rho}\left( t\right) =\left[ 
\begin{array}{cc}
\text{cos}^{2}\left( \tfrac{\beta }{2}\right)  & \frac{1}{2}\text{sin}\left(
\beta \right) e^{-2t\left( iB+\gamma \right) } \\ 
\frac{1}{2}\text{sin}\left( \beta \right) e^{-2t\left( \gamma -iB\right) } & 
\text{sin}^{2}\left( \tfrac{\beta }{2}\right) 
\end{array}%
\right] .
\end{equation}%
The basis-independent expression of QFI for a two-level mixed state $\hat{%
	\rho}\left( t\right) $ is of the following form \cite{liu2020quantum}
\begin{equation}
F_{\theta }\left[ \hat{\rho}\left( t\right) \right] =\text{Tr}\left[ \left(
\partial _{\theta }\hat{\rho}\left( t\right) \right) ^{2}\right] +\frac{%
	\text{Tr}\left[ \left( \hat{\rho}\left( t\right) \partial _{\theta }\hat{\rho%
	}\left( t\right) \right) ^{2}\right] }{\text{det}\left[ \hat{\rho}\left(
	t\right) \right] },
\label{Eq4s}
\end{equation}%
where $\theta $ refers to the parameter to be estimated. When $\theta $ is taken as  $\beta $ and $B$, respectively, we obtain 
\begin{subequations}
	\begin{eqnarray}
	F_{\beta }\left[ \hat{\rho}\left( t\right) \right]  &=&1, \\
	F_{B}\left[ \hat{\rho}\left( t\right) \right]  &=&4\text{sin}^{2}\left(
	\beta \right) e^{-4\gamma t}t^{2}.
	\end{eqnarray}%
\end{subequations}
This is Eqs.~8(a) and 8(b) in the main text.

\section{IV. Affine transformation of the Bloch sphere}
To gain a visual understanding of the impact of noise on the dynamics and subsequently on the QFI, we employ the affine transformation to map the master equation corresponding to Scenario 2 in the main text onto the Bloch sphere.
When $\vartheta =\pi/2$ (the magnetic field direction is along the $Z$-axis) and $\alpha \neq k\pi /2$ $(k\in \mathbb{Z})$, $\hat{H}=B\hat{\sigma}_{z}$ and jump operator $\hat{\Gamma}=\cos (\alpha )\hat{\sigma}_{x}+\sin (\alpha )\hat{\sigma}_{z}$, indicating NPC noise. 
The specific form of the master equation is as follows,
\begin{equation}
\frac{d\hat{\rho}\left( t\right) }{dt}=( \hat{\mathcal{H}}+\hat{%
	\mathcal{L}}) \hat{\rho}\left( t\right) =\mathbf{\hat{\Lambda}}[%
\hat{\rho}\left( t\right) ],  \label{es35}
\end{equation}%
with 
\begin{eqnarray}
\mathbf{\hat{\Lambda}}[\hat{\rho}\left( t\right) ] &=&-i[B\hat{\sigma}%
_{z},\hat{\rho}\left( t\right) ]+S_{1}\left[ \hat{\sigma}_{z}\hat{\rho}%
\left( t\right) \hat{\sigma}_{z}-\hat{\rho}\left( t\right) \right] +
\notag \\
&&S_{2}\left[ \hat{\sigma}_{+}\hat{\rho}\left( t\right) \hat{\sigma}_{+}+%
\hat{\sigma}_{-}\hat{\rho}\left( t\right) \hat{\sigma}_{-}\right] + 
\notag \\
&&S_{2}\left[ \hat{\sigma}_{+}\hat{\rho}\left( t\right) \hat{\sigma}_{-}-%
\frac{1}{2}\hat{\sigma}_{-}\hat{\sigma}_{+}\hat{\rho}\left( t\right) -%
\frac{1}{2}\hat{\rho}\left( t\right) \hat{\sigma}_{-}\hat{\sigma}_{+}%
\right] +  \notag \\
&&S_{2}\left[ \hat{\sigma}_{-}\hat{\rho}\left( t\right) \hat{\sigma}_{+}-%
\frac{1}{2}\hat{\sigma}_{+}\hat{\sigma}_{-}\hat{\rho}\left( t\right) -%
\frac{1}{2}\hat{\rho}\left( t\right) \hat{\sigma}_{+}\hat{\sigma}_{-}%
\right] +  \notag \\
&&S_{3}\left[ \hat{\sigma}_{+}\hat{\rho}\left( t\right) \hat{\sigma}_{z}-%
\frac{1}{2}\hat{\sigma}_{z}\hat{\sigma}_{+}\hat{\rho}\left( t\right) -%
\frac{1}{2}\hat{\rho}\left( t\right) \hat{\sigma}_{z}\hat{\sigma}_{+}%
\right] +  \notag \\
&&S_{3}\left[ \hat{\sigma}_{-}\hat{\rho}\left( t\right) \hat{\sigma}_{z}-%
\frac{1}{2}\hat{\sigma}_{z}\hat{\sigma}_{-}\hat{\rho}\left( t\right) -%
\frac{1}{2}\hat{\rho}\left( t\right) \hat{\sigma}_{z}\hat{\sigma}_{-}%
\right] +  \notag \\
&&S_{3}\left[ \hat{\sigma}_{z}\hat{\rho}\left( t\right) \hat{\sigma}_{+}-%
\frac{1}{2}\hat{\sigma}_{+}\hat{\sigma}_{z}\hat{\rho}\left( t\right) -%
\frac{1}{2}\hat{\rho}\left( t\right) \hat{\sigma}_{+}\hat{\sigma}_{z}%
\right] +  \notag \\
&&S_{3}\left[ \hat{\sigma}_{z}\hat{\rho}\left( t\right) \hat{\sigma}_{-}-%
\frac{1}{2}\hat{\sigma}_{-}\hat{\sigma}_{z}\hat{\rho}\left( t\right) -%
\frac{1}{2}\hat{\rho}\left( t\right) \hat{\sigma}_{-}\hat{\sigma}_{z}%
\right] ,  \label{es36}
\end{eqnarray}
where $S_{1}=\gamma \sin ^{2}\left( \alpha \right) ,S_{2}=\gamma \cos
^{2}\left( \alpha \right) $ and $S_{3}=\dfrac{\gamma }{2}\sin (2\alpha )$. Also, $\hat{\sigma}_{+}=\left\vert e\right\rangle \left\langle g\right\vert $ and $%
\hat{\sigma}_{-}=$ $\left\vert g\right\rangle \left\langle e\right\vert $
denote respectively the flip-up and flip-down operators, satisfying $\hat{\sigma}_{x}=$ 
$\hat{\sigma}_{+}+\hat{\sigma}_{-}$.

Let $\{ \mathbf{%
	\hat{G}}_{i}\} $ $\left( i=0,1,2,3\right) $ denotes four basis matrices we have for
the Bloch sphere~\cite{PCNPC}, 
\begin{equation}
\{ \mathbf{\hat{G}}_{i}\} _{i=0,1,2,3}=\frac{1}{\sqrt{2}}\left\{ 
\mathbf{I}_{2},\hat{\sigma}_{x},\hat{\sigma}_{y},\hat{\sigma}_{z}\right\} .
\end{equation}%
We can then represent $\hat{\rho}\left( t\right)$ and $\mathbf{\hat{\Lambda}}[\hat{\rho}\left( t\right)]$ using the basis set ${ \{\mathbf{\hat{G}}_{i}}\}$,  
\begin{eqnarray}
\hat{\rho}\left( t\right)  &=&\sum_{i=0}^{3}x_{i}\mathbf{\hat{G}}_{i}=%
\frac{1}{2}\left[ \mathbf{I}_{2}+\vec{r}(t)\cdot \hat{\sigma}\right] ,  \label{es38}
\\
\mathbf{\hat{\Lambda}}[\hat{\rho}\left( t\right) ]
&=&\sum_{i,j=0}^{3}\Lambda _{ij}x_{j}\mathbf{\hat{G}}_{i},
\end{eqnarray}%
with%
\begin{equation}
x_{i}=\text{Tr}\left[ \mathbf{\hat{G}}_{i}^{\dag }\hat{\rho}\left(
t\right) \right] ,\Lambda _{ij}=\text{Tr}\left[ \mathbf{\hat{G}}_{i}^{\dag }%
\mathbf{\hat{\Lambda}}[\mathbf{\hat{G}}_{j}]\right] , \label{es43}
\end{equation}%
where $x_{i}$ is the component of $\hat{\rho}\left( t\right) $ with
respect to $\mathbf{\hat{G}}_{i}$, $\mathbf{\hat{\sigma}}$ is vector of Pauli matrices, and $\vec{r}(t)$ is the Bloch vector. Further, the master Eq.~(\ref{es35}) can be reduced to the following matrix form, 
\begin{equation}
\label{Eqs39}
\frac{d\vec{x}}{dt}=\Lambda \vec{x},
\end{equation}%
where $\vec{x}$ is $4\times 1$ dimension column vector and and the coefficient matrix $\Lambda$ is given by 
\begin{equation}
\Lambda =\left[ 
\begin{array}{cccc}
0 & 0 & 0 & 0 \\ 
0 & -2S_{1} & -2B & 2S_{3} \\ 
0 & 2B & -2(S_{1}+S_{2}) & 0 \\ 
0 & 2S_{3} & 0 & -2S_{2}%
\end{array}%
\right] .
\label{Eqs30}
\end{equation}%

Notice that, we need not consider all elements of $\vec{x}(t)$ since the calculation of QFI depends solely on the Bloch vector $\vec{r}(t)=\left[ \vec{x}\left( 2,1\right) ,\vec{x}\left( 3,1\right) ,%
\vec{x}\left( 4,1\right) \right] ^T$. Thus, from Eq.~(\ref{Eqs39}) and (\ref{Eqs30}) we can derive $\vec{r}(t)=e^{Dt}%
\vec{r}\left( 0\right) =\Upsilon_{\text{npc}} \left( t\right) \vec{r}\left( 0\right) $, with 
\begin{equation}
D=\left[ 
\begin{array}{ccc}
-2S_{1} & -2B & 2S_{3} \\ 
2B & -2(S_{1}+S_{2}) & 0 \\ 
2S_{3} & 0 & -2S_{2}%
\end{array}%
\right]~.
\end{equation}%
The exact form of the affine transformation matrix $\Upsilon_{\text{npc}} \left( t\right) $ is very complex,
so we adopt a short-time approximation 
\begin{eqnarray}
\Upsilon_{\text{npc}} \left( t\right)  &=&e^{Dt}=\sum_{k=0}^{\infty }\frac{\left(
	Lt\right) ^{k}}{k!}=\mathbf{I}_{3}+Dt+\frac{D^{2}t^{2}}{2}+\frac{D^{3}t^{3}}{3!}%
+o(t^{4})  \notag \\
&=&\left[ 
\begin{array}{ccc}
\Upsilon _{11} & \Upsilon _{12} & \Upsilon _{13} \\ 
\Upsilon _{21} & \Upsilon _{22} & \Upsilon _{23} \\ 
\Upsilon _{31} & \Upsilon _{32} & \Upsilon _{33}%
\end{array}%
\right] , \label{es47}
\end{eqnarray}%
where
\begin{eqnarray*}
	\Upsilon _{11} &=&1-2S_{1}t+2\left( S_{3}^{2}+S_{1}^{2}-B^{2}\right)
	t^{2}+\left( 4B^{2}S_{1}+\frac{4}{3}B^{2}S_{2}-\frac{8}{3}S_{3}^{2}S_{1}-%
	\frac{4}{3}S_{3}^{2}S_{2}-\frac{4}{3}S_{1}^{3}\right) t^{3}, \\
	\Upsilon _{12} &=&-\frac{2}{3}Bt\left(
	3-6S_{1}t-3S_{2}t-2B^{2}t^{2}+2S_{3}^{2}t^{2}+6S_{1}^{2}t^{2}+2S_{2}^{2}t^{2}+6S_{1}S_{2}t^{2}\right) ,
	\\
	\Upsilon _{13} &=&\frac{2}{3}S_{3}t\left(
	3-3S_{1}t-3S_{2}t-2B^{2}t^{2}+2S_{3}^{2}t^{2}+2S_{1}^{2}t^{2}+2S_{1}S_{2}t^{2}+2S_{2}^{2}t^{2}\right) ,
	\\
	\Upsilon _{21} &=&\frac{2}{3}Bt\left(
	3-6S_{1}t-3S_{2}t-2B^{2}t^{2}+2S_{3}^{2}t^{2}+6S_{1}^{2}t^{2}+6S_{1}S_{2}t^{2}+2S_{2}^{2}t^{2}\right) ,
	\\
	\Upsilon _{22} &=&1-2\left( S_{1}+S_{2}\right) t+2\left[ \left(
	S_{1}+S_{2}\right) ^{2}-B^{2}\right] t^{2}+\frac{1}{3}\left[ 4B^{2}\left(
	2S_{1}+S_{2}\right) -4\left( S_{1}+S_{2}\right) \left( \left[ \left(
	S_{1}+S_{2}\right) ^{2}-B^{2}\right] \right) \right] t^{3}, \\
	\Upsilon _{23} &=&-\frac{2}{3}BS_{3}t^{2}\left( 4S_{1}t+4S_{2}t-3\right) , \\
	\Upsilon _{31} &=&\frac{2}{3}S_{3}t\left(
	3-3S_{1}t-3S_{2}t-2B^{2}t^{2}+2S_{3}^{2}t^{2}+2S_{1}^{2}t^{2}+2S_{1}S_{2}t^{2}+2S_{2}^{2}t^{2}\right) ,
	\\
	\Upsilon _{32} &=&\frac{2}{3}BS_{3}t^{2}\left( 4S_{1}t+4S_{2}t-3\right) , \\
	\Upsilon _{33} &=&1-2S_{2}t+2\left( S_{3}^{2}+S_{2}^{2}\right) t^{2}-\frac{1%
	}{3}\left[ 4S_{3}^{2}\left( S_{1}+2S_{2}\right) +4S_{2}^{3}\right] t^{3}.
\end{eqnarray*}%

The singular value decomposition of $\Upsilon_{\text{npc}} (t)$ satisfies the following form~\cite{PCNPC}
\begin{equation}
\Upsilon_{\text{npc}} (t)=R_{m1}^{\varphi _{1}}V_{\text{npc}}R_{m2}^{\varphi_{2}},
\end{equation}%
where $V_{\text{npc}}=\text{diag}\left\{ n_{x},n_{y},n_{z}\right\} $ denotes a diagonal matrix, representing the unequal contractions of the Bloch sphere along the $X$, $Y$ and $Z$ axes, $R_{m_{i}}^{\varphi _{i}}(i=1,2)$ represents a rotation of the Bloch spere around the $m_{i}$ axis by an angle $\varphi _{i}$. Especially, in this case, the contraction and rotation matrices are non-commutative.

Specifically, suppose the initial state of system is $\left\vert \Phi \left( 0\right) \right\rangle =\left\vert
g\right\rangle $, the corresponding Bloch vector is given by $\vec{r}\left( 0\right) =\left[ 0,0,-1\right] ^{\text{T}}$ and lies on the negative Z-axis. 
In the short time 
\begin{equation}
\vec{r}\left( t\right) =\Upsilon_{\text{npc}} \left( t\right) \vec{r}\left( 0\right) =-%
\left[ \Upsilon _{13},\Upsilon _{23},\Upsilon _{33}\right] ^{\text{T}},
\label{es48}
\end{equation}%
which is Eq.~(9) in the main text.  Based on this result, we can clearly see that  $\vec{r}\left( 0\right) $ deviate from the $Z$
axis under the action of $\Upsilon (t)$, and get the quantum coherence.  

With the Bloch vector, the calculation formula of QFI can be written as~\cite{liu2020quantum} 
\begin{equation}
F_{B}(t) =\left\vert \partial _{B}%
\vec{r}\left( t\right) \right\vert ^{2}+\frac{\left[ \vec{r}\left( t\right)
	\cdot \partial _{B}\vec{r}\left( t\right) \right] ^{2}}{1-\left\vert \vec{r}%
	\left( t\right) \right\vert ^{2}}.  \label{es49}
\end{equation}%
By combining Eqs.~(\ref{es48}) and (\ref{es49}), it becomes evident that $F_{B}(t)\neq 0$ due to the presence of the parameter $B$ in $\Upsilon_{13}$ and $\Upsilon _{23}$ when $S_{3}\neq 0$ (i.e., $\alpha \neq k\pi/2,k\in \mathbb{Z}$). 
Notably, in the absence of noise, specifically when $\gamma=0$, we have  $F_{B}(t)=0$.
Hence, we demonstrate that the introduction of NPC noise facilitates a qualitative transition in the QFI, increasing from nothing.
It is crucial to acknowledge that not all NPC noise is capable of enhancing quantum metrology. 
For example, the case of $\hat{\Gamma}=\hat{\sigma}_{x}$ corresponds to the NPC noise, resulting in $S_{2}\neq 0$ and  $S_{3}= 0$. However, under these conditions, we have  $\Upsilon _{13}= 0$ and $\Upsilon_{23}=0$, leading to $F_{B}(t)= 0$. Therefore, only the NPC noise capable of inducing nonzero $S_3$ terms in the master equation (\ref{es36}) can enhance QFI, as the $S_3$ terms represent the correlation between the dissipation channels.

Instead, for the PC noise, the affine transformation matrix $\Upsilon _{%
	\text{pc}} $ can always be
written in the following general form~\cite{PCNPC}, 
\begin{equation}
\Upsilon _{\text{pc}}(t)=\left[ 
\begin{array}{ccc}
n_{\bot }\cos \left( \xi \right)  & -n_{\bot }\sin \left( \xi \right)  & 0
\\ 
n_{\bot }\sin \left( \xi \right)  & n_{\bot }\cos \left( \xi \right)  & 0 \\ 
0 & 0 & n_{\Vert }%
\end{array}%
\right] ,
\end{equation}%
and its singular value decomposition is 
\begin{equation}
\Upsilon _{\text{pc}}(t)=R_{Z}^{\xi }V_{\text{pc}},
\end{equation}%
with
\begin{equation}
R_{Z}^{\xi }=\left[ 
\begin{array}{ccc}
\cos \left( \xi \right)  & -\sin \left( \xi \right)  & 0 \\ 
\sin \left( \xi \right)  & \cos \left( \xi \right)  & 0 \\ 
0 & 0 & 1%
\end{array}%
\right] ,V_{\text{pc}}=\left[ 
\begin{array}{ccc}
n_{\bot } & 0 & 0 \\ 
0 & n_{\bot } & 0 \\ 
0 & 0 & n_{\Vert }%
\end{array}%
\right] .
\end{equation}%
Here, $V_{\text{pc}}$ makes the Bloch sphere equal contractions along the $X$
and $Y$ axes with amplitude $n_{\bot }$, while flattening it along the $Z$
axis by an amplitude $n_{\Vert }$; $R_{Z}^{\xi }$ describes the Bloch sphere
rotate angle $\xi $ around the $Z$-axis. Note that in the current case the contraction matrix $%
V_{\text{pc}}$ commutes with the rotation matrix $R_{Z}^{\xi }$, namely $[
V_{\text{pc}},R_{Z}^{\xi }] =0$. Consequently, $\vec{r}(t)$ remains aligned with the Z-axis, and is unable to obtain quantum coherence under the action of $\Upsilon _{\text{pc}}$, 
Therefore, it is impossible to extract information about $B$ from $\vec{r}\left( t\right) $.

\section{V. Effective Hamiltonian theory}
In order to more conveniently demonstrate the relationship between QFI and noise in spin-1/2 systems, we adopt the effective Hamiltonian method, which can obtain an effective Hamiltonian that can describe the main dynamics of the system in a noisy environment.

The total Hamiltonian including system-environment coupling writes
$\left( \hslash =1\right) $
\begin{eqnarray}
\hat{H}_{\text{SE}} &=&\hat{H}_{\text{S}}+\hat{H}_{\text{E}}+\hat{H}_{\text{%
		I}}  \notag \\
&=&B\left[ \text{cos}\left( \vartheta \right) \hat{\sigma}_{x}+\text{sin}%
\left( \vartheta \right) \hat{\sigma}_{z}\right] +\sum_{k}\omega
_{k}b_{k}^{\dag }b_{k}+\hat{\Gamma}\otimes \sum_{k}g_{k}( b_{k}^{\dag }+b_{k}) ,
\end{eqnarray}%
where $b_{k}( b_{k}^{\dag }) $ is annihilation (creation)
operator of the $k$th bath mode with frequency $\omega _{k}$, $g_{k}$ denotes the coupling strength between the two-level system and the $k$th bath
mode, $\hat{\Gamma}=$ cos$\left( \alpha \right) \hat{\sigma}_{x}+$sin$\left(
\alpha \right) \hat{\sigma}_{z}$ is jump operator. 
The coupling between the system and the bath is fully captured by the
spectral density function, i.e., $J\left( \omega \right) =\sum_{k}g_{k}^{2}\delta
\left( \omega -\omega _{k}\right)$.

Defining the coupling parameter $g_{\text{r}}$ between the system and the reaction coordinate, along with the frequency of the reaction coordinate $\Omega_{\text{r}}$, as follows~\cite{PRXZ}:
\begin{eqnarray}
g_{\text{r}}^2 &=&\frac{1}{\Omega _{\text{r}}}\int_{0}^{\infty }\omega J\left(
\omega \right) d\omega , \\
\Omega _{\text{r}}^2 &=&\frac{\int_{0}^{\infty }\omega ^{3}J\left( \omega
	\right) d\omega }{\int_{0}^{\infty }\omega J\left( \omega \right) d\omega }.
\end{eqnarray}%
When the condition $g_{r}/\Omega_{r} \ll 1$ is satisfied, we can use the ``Reaction-Coordinate Polaron-Transform (RCPT) Framework'' method proposed by Nicholas Anto-Sztrikacs et al.~\cite{PRXZ} to derive an effective Hamiltonian that accurately captures the dominant dynamics of the system within the open environment. The expression for this effective Hamiltonian is given by
\begin{eqnarray}
\label{Eqs94}
\hat{H}_{\text{S}}^{\text{eff}} &=&e^{-\left( g_{\text{r}}^{2}/2\Omega _{%
		\text{r}}^{2}\right) \hat{\Gamma}^{2}}\left( \sum_{n=0}^{\infty }\frac{g_{%
		\text{r}}^{2n}}{\Omega _{\text{r}}^{2n}n!}\hat{\Gamma}^{n}\hat{H}_{\text{S}}%
\hat{\Gamma}^{n}\right) e^{-\left( g_{\text{r}}^{2}/2\Omega _{\text{r}%
	}^{2}\right) \hat{\Gamma}^{2}}  \notag \\
&=&e^{-\left( g_{\text{r}}^{2}/2\Omega _{\text{r}}^{2}\right) }\left[
\sum_{n=0}^{\infty }\ \frac{\left( g_{\text{r}}^{2}/\Omega _{\text{r}%
	}^{2}\right) ^{2n}}{2n!}\hat{H}_{\text{S}}+\sum_{n=0}^{\infty }\ \frac{%
	\left( g_{\text{r}}^{2}/\Omega _{\text{r}}^{2}\right) ^{2n+1}}{\left(
	2n+1\right) !}\hat{\Gamma}\hat{H}_{\text{S}}\hat{\Gamma}\right] e^{-\left(
	g_{\text{r}}^{2}/2\Omega _{\text{r}}^{2}\right) }  \notag \\
&=&\frac{1}{2}\zeta _{+}\hat{H}_{\text{S}}+\frac{1}{2}\zeta _{-}\hat{\Gamma}%
\hat{H}_{\text{S}}\hat{\Gamma},
\end{eqnarray}%
where $\zeta _{\pm }=\frac{1}{2}\left[ 1\pm e^{-2g_{\text{r}}^{2}/\Omega _{\text{r}%
	}^{2}}\right]$, and we used equations $\hat{\Gamma}^{2n}={\bf{I}}_2$ and $\hat{\Gamma}^{2n+1}=\hat{\Gamma}$.

When $\alpha \neq \vartheta + 2k\pi$ (where $k$ is an integer), we encounter NPC noise, and the corresponding effective Hamiltonian is given by
\begin{equation}
\label{Eqs98}
\hat{H}_{\text{S}}^{\text{eff}}=B\left[ \zeta _{+}\text{cos}\left( \vartheta
\right) +\zeta _{-}\text{cos}\left( 2\alpha -\vartheta \right) \right] \hat{%
	\sigma}_{x}+B\left[ \zeta _{+}\text{sin}\left( \vartheta \right) +\zeta _{-}%
\text{sin}\left( 2\alpha -\vartheta \right) \right] \hat{\sigma}_{z}.
\end{equation}
By comparing $\hat{H}_\text{S}$ and $\hat{H}_\text{S}^{\text{eff}}$, we can observe that the environmental coupling introduces modulation in spin flipping and spin splitting that depends on the difference between $\vartheta$ and $\alpha$.

Utilizing the method developed by Wang et al. \cite{PRAJXX}, it has been established that when the Hamiltonian takes the form $\hat{H}=\vec{N}\left( \vartheta \right) \cdot \vec{J}$, where  $\vec{J}=[\hat{\sigma}_{x}/2,0,\hat{\sigma}_{z}/2]$. The corresponding maximum QFI is given by
\begin{equation}
\label{S44}
F_{\vartheta}^{\text{max}}\left(t\right) = \left( \frac{d|\vec{N}|}{%
	d\vartheta }\vec{e}_{N}\right) ^{2}t^{2}+4\left( \frac{d\vec{e}_{N}}{%
	d\vartheta }\right) ^{2}\text{sin}^{2}\left( \frac{|\vec{N}|}{2}t\right).
\end{equation}
This expression consists of two parts: one that increases quadratically with time, attributed to the dependence of $\vec{N}$'s magnitude on $\vartheta$, and the other that oscillates with time, attributed to the dependence of $\vec{N}$'s direction on $\vartheta$.

$\hat{H}_\text{S}^{\text{eff}}$ can be rewritten in the form of a vector dot product as follows:
\begin{equation}
\hat{H}_{\text{S}}^{\text{eff}}=\vec{R}^{\prime }\cdot \vec{J},
\end{equation}
with
\begin{eqnarray}
\vec{R}^{\prime } &=&\left\{ 2B\left[ \zeta _{+}\text{cos}\left(
\vartheta \right) +\zeta _{-}\text{cos}\left( 2\alpha -\vartheta \right) %
\right] ,0,B\left[ \zeta _{+}\text{sin}\left( \vartheta \right) +\zeta _{-}%
\text{sin}\left( 2\alpha -\vartheta \right) \right] \right\} =|\vec{R}^{\prime }|\vec{e}_{^{R^{\prime }}} \\
|\vec{R}^{\prime}| &=& 2B\sqrt{\zeta _{+}^{2}+\zeta _{-}^{2}+2\zeta _{+}\zeta _{-}\text{cos}%
	\left( 2\alpha -2\vartheta \right) }, \\
\vec{e}_{R^{\prime }} &=&\left[ \frac{\zeta _{+}\text{cos}\left(
	\vartheta \right) +\zeta _{-}\text{cos}\left( 2\alpha -\vartheta \right) }{%
	\sqrt{\zeta _{+}^{2}+\zeta _{-}^{2}+2\zeta _{+}\zeta _{-}\text{cos}\left(
		2\alpha -2\vartheta \right) }},0,\frac{\zeta _{+}\text{sin}\left( \vartheta
	\right) +\zeta _{-}\text{sin}\left( 2\alpha -\vartheta \right) }{\sqrt{\zeta
		_{+}^{2}+\zeta _{-}^{2}+2\zeta _{+}\zeta _{-}\text{cos}\left( 2\alpha
		-2\vartheta \right) }}\right], \\
\end{eqnarray}%
where $|\vec{R}^{\prime}|$ and $\vec{e}_{R^{\prime }}$ are the magnitude and unit vector of $\vec{R}^{\prime }$, respectively.

For comparison, in the absence of noise, we have 
\begin{equation}
\hat{H}_{\text{S}}=\vec{R}\cdot \vec{J},
\end{equation}
where
\begin{equation}
\vec{R}=\left[ 2B\text{cos}\left( \vartheta \right) ,0,2B\text{sin}\left(
\vartheta \right) \right] .
\end{equation}
We observe that $\vec{R}$'s direction depends on $\vartheta$ but its magnitude is constant. Conversely,  
$\vec{R}^\prime$'s direction and magnitude both depend on $\vartheta$, i.e.,
\begin{equation}
\frac{d|\vec{R}|}{d\vartheta }=0,\frac{d|\vec{R}^{\prime }|}{d\vartheta }%
\neq 0.
\end{equation}
Therefore, in the absence of noise, the maximum QFI regarding $\vartheta$ is given by
\begin{equation}
F_{\vartheta} = 4\sin^2(Bt),
\end{equation}
with an extreme value of $4$. In contrast, the effective Hamiltonian $\hat{H}_{\text{S}}^{\text{eff}}$ governing the dynamics in the presence of NPC noise leads to a non-zero first term on the right-hand side of Eq.~(\ref{S44}). This modulation of the QFI scales with the square of the evolution time, indicating a rapid increase compared to the noiseless case.

\section{VI. Timescale for an open system to reach steady state}

The dynamics of an open system is characterized by a master equation within a Hilbert space,
\begin{eqnarray}
\label{Eq22s}
\frac{d\hat{\rho}\left( t\right) }{dt} &=&-i[ \hat{H}_{\text{S}},\hat{\rho}\left(
t\right) ] +\gamma \lbrack \hat{\Gamma}\hat{\rho}\left( t\right) \hat{%
	\Gamma}^{\dag }-\frac{1}{2}\hat{\Gamma}^{\dag }\hat{\Gamma}\hat{\rho}\left( t\right) -%
\frac{1}{2}\hat{\rho}\left( t\right) \hat{\Gamma}^{\dag }\hat{\Gamma}],
\end{eqnarray}%
but it is often more convenient to examine the timescale at which an open system attains steady state within the Liouville space where the density matrix is vectorized and Eq.~(\ref{Eq22s}) transforms into~\cite{alipour2014quantum}
\begin{eqnarray}
\frac{d|\hat{\rho}\left( t\right) \rangle \rangle }{dt} &=&(\tilde{L}_{1}+%
\tilde{L}_{2})|\hat{\rho}\left( t\right) \rangle \rangle=\tilde{L}_{\text{tot}}|\hat{\rho}\left( t\right) \rangle \rangle ,
\end{eqnarray}%
with
\begin{eqnarray}
\tilde{L}_{1} &=&-i( \hat{H}_{\text{S}}\otimes {\bf{I}}_2-{\bf{I}}_2\otimes \hat{H}_{\text{S}}) , \\
\tilde{L}_{2} &=&\gamma \left( \hat{\Gamma}\otimes \hat{\Gamma}^{\ast }-%
\frac{1}{2}\hat{\Gamma}^{\dag }\hat{\Gamma}\otimes {\bf{I}}_2-\frac{1}{2}{\bf{I}}_2\otimes 
\hat{\Gamma}^{\text{T}}\hat{\Gamma}^{\ast }\right) ,
\end{eqnarray}%
where $|\hat{\rho}\left( t\right) \rangle \rangle $ refers to the column
vector form of the density matrix $\hat{\rho}\left( t\right) $. 
The Liouville supermatrix, denoted as $\tilde{L}_{%
	\text{tot}}$, satisfies the eigenequation
\begin{equation}
\tilde{L}_{\text{tot}}|\tilde{L}_{\text{tot}}^{i}\rangle \rangle =\tilde{L}_{%
	\text{tot}}^{i}|\tilde{L}_{\text{tot}}^{i}\rangle \rangle ,
\end{equation}%
where $\tilde{L}_{\text{tot}}^{i}$ and $|\tilde{L}_{\text{tot}}^{i}\rangle
\rangle $ $\left( i=1\sim4\right) $ are the eigenvalues and eigenvectors of $%
\tilde{L}_{\text{tot}}$, respectively. Without losing generality, we will
sort $\tilde{L}_{\text{tot}}^{i}$ in the following way, i.e., Re$[ 
\tilde{L}_{\text{tot}}^{1}] =0\geq $ Re$[ \tilde{L}_{\text{tot}%
}^{2}] \geq $ Re$[ \tilde{L}_{\text{tot}}^{3}] \geq $ Re$%
[ \tilde{L}_{\text{tot}}^{4}] $~\cite{rivas2012open,breuer2002theory}. 
The steady state is governed by the eigenvectors of $\tilde{L}_{\text{tot}}$ that are associated with a zero eigenvalue. The timescale to approach this steady state is influenced by the Liouvillian eigenvalue whose real part is closest to zero, and can be estimated using the inverse of its real part.

In our case $\hat{H}_{\text{S}}=B\left[ \text{cos}\left( \theta \right) \hat{\sigma}_{x}+%
\text{sin}\left( \theta \right) \hat{\sigma}_{z}\right] $ and $\hat{\Gamma}=$
cos$\left( \alpha \right) \hat{\sigma}_{x}+$sin$\left( \alpha \right) \hat{%
	\sigma}_{z}$. 
When $\theta =\alpha $, the quantum jump operator $\hat{\Gamma}$ induced
generalized dephasing noise. At this moment, $\tilde{L}_{\text{tot}}^{1}=%
\tilde{L}_{\text{tot}}^{2}=0$, indicating the occurrence of degeneracy, and
the corresponding eigenvectors are 
\begin{eqnarray}
|\tilde{L}_{\text{tot}}^{1}\rangle \rangle &\propto &\left[ 
\begin{array}{c}
\text{cos}^{2}\left( \frac{\pi }{4}-\frac{\theta }{2}\right) \\ 
\frac{1}{2}\text{cos}\left( \theta \right) \\ 
\frac{1}{2}\text{cos}\left( \theta \right) \\ 
\text{sin}^{2}\left( \frac{\pi }{4}-\frac{\theta }{2}\right)%
\end{array}%
\right] , \\
|\tilde{L}_{\text{tot}}^{2}\rangle \rangle &\propto &\left[ 
\begin{array}{c}
\text{sin}^{2}\left( \frac{\pi }{4}-\frac{\theta }{2}\right) \\ 
-\frac{1}{2}\text{cos}\left( \theta \right) \\ 
-\frac{1}{2}\text{cos}\left( \theta \right) \\ 
\text{cos}^{2}\left( \frac{\pi }{4}-\frac{\theta }{2}\right)%
\end{array}%
\right] .
\end{eqnarray}%
The steady-state density matrix $\hat{\rho}_{\text{sts}}$ can be obtain by projecting them back onto the Hilbert space,
\begin{eqnarray}
|\tilde{L}_{\text{tot}}^{1}\rangle \rangle &\rightarrow &\hat{\rho}_{\text{%
		sts}}^{1}=\left[ 
\begin{array}{cc}
\text{cos}^{2}\left( \frac{\pi }{4}-\frac{\theta }{2}\right) & \frac{1}{2}%
\text{cos}\left( \theta \right) \\ 
\frac{1}{2}\text{cos}\left( \theta \right) & \text{sin}^{2}\left( \frac{\pi 
}{4}-\frac{\theta }{2}\right)%
\end{array}%
\right] , \\
|\tilde{L}_{\text{tot}}^{2}\rangle \rangle &\rightarrow &\hat{\rho}_{\text{%
		sts}}^{2}=\left[ 
\begin{array}{cc}
\text{sin}^{2}\left( \frac{\pi }{4}-\frac{\theta }{2}\right) & -\frac{1}{2}%
\text{cos}\left( \theta \right) \\ 
-\frac{1}{2}\text{cos}\left( \theta \right) & \text{cos}^{2}\left( \frac{\pi 
}{4}-\frac{\theta }{2}\right)%
\end{array}%
\right] ,
\end{eqnarray}%
and
\begin{equation}
\hat{\rho}_{\text{sts}}=P_{1}\hat{\rho}_{\text{sts}}^{1}+P_{2}\hat{\rho}_{%
	\text{sts}}^{2},
\end{equation}%
where the probabilities $P_{1}$ and $P_{2}$ depend on the initial state of the system. The timescale for the system to reach the steady state is~\cite{rivas2012open,breuer2002theory,zhou2023accelerating} 
\begin{equation}
t_{\text{sts}}\sim \frac{1}{\left\vert \text{Re}[ \tilde{L}_{\text{tot}%
	}^{3}] \right\vert }.
\end{equation}%

When $\theta \neq \alpha $, the noise induced by the quantum jump operator $\hat{\Gamma}$ is no longer purely dephasing. At this moment, only $\tilde{L}_{\text{tot}%
}^{1}=0$, and the corresponding eigenvector is 
\begin{equation}
|\tilde{L}_{\text{tot}}^{1}\rangle \rangle \propto \left[ 
\begin{array}{c}
\frac{1}{2} \\ 
0 \\ 
0 \\ 
\frac{1}{2}%
\end{array}%
\right] .
\end{equation}%
So the steady-state density matrix is
\begin{equation}
|\tilde{L}_{\text{tot}}^{1}\rangle \rangle \rightarrow \hat{\rho}_{\text{sts}%
}=\left[ 
\begin{array}{cc}
\frac{1}{2} & 0 \\ 
0 & \frac{1}{2}%
\end{array}%
\right] ,
\end{equation}%
which, different from the previous case, is unique and
does not depend on the initial state of the system. The timescale for reaching the steady state is~\cite{rivas2012open,breuer2002theory,zhou2023accelerating} 
\begin{equation}
t_{\text{sts}}\sim \frac{1}{\left\vert \text{Re}[ \tilde{L}_{\text{tot}%
	}^{2}] \right\vert }.
\end{equation}%
Comparing the two situations above, it becomes evident that $\theta =\alpha $ serves as a critical point for noise properties. Whether $\theta $ and $\alpha 
$ are equal determines the emergence of two distinctly different steady states.
When $\theta $ and $%
\alpha $ are very close, it can be observed that Re$[ \tilde{L}_{\text{tot}}^{2}]$ approaches zero, resulting in a significantly prolonged time scale for the system to reach a steady state ($t_{\text{sts}}\rightarrow \infty$). This also means that the available time for system coding  parameter has increased significantly. Consequently, this provides a physical explanation for why the introduction of NPC-noise is more prone to overcoming the precision limit established by coherent dynamics when $\theta $ and $%
\alpha $ are closely aligned, as shown in the main text.

\section{VII. Multi-particle situation}
We will demonstrate in this section that the results obtained for the single-particle case in the main text can be extended to the multi-particle case.
For an open collective system consisting of $N$
spin-1/2 particles in a magnetic field,
its dynamics is described by quantum master equation
\begin{equation}
\label{EqS9}
\frac{d\hat{\rho}_{\text{tot}}\left( t\right) }{dt}-i[\hat{H}_{\text{tot}%
},\hat{\rho}_{\text{tot}}\left( t\right) ]+\sum_{i=1}^{N}\gamma _{i}[\hat{%
	\Gamma}_{i}\hat{\rho}_{\text{tot}}\left( t\right) \hat{\Gamma}_{i}^{\dag } 
-\frac{1}{2}\{\hat{\Gamma}_{i}^{\dag }\hat{\Gamma}_{i},\hat{\rho}_{\text{%
		tot}}\left( t\right) \}],
\end{equation}
where $\hat{H}_{\text{tot}}=\sum_{i=1}^{N}B\left[ \cos (\vartheta )\hat{\sigma}%
_{x}^{i}+\sin (\vartheta )\hat{\sigma}_{z}^{i}\right] $ is the total Hamiltonian of system. $\hat{\Gamma}_{i}=\cos (\alpha)\hat{\sigma}_{x}^{i}+\sin (\alpha)\hat{\sigma}_{z}^{i}$ is the quantum jump operator acting
on the $i$th particle, and the corresponding decay rate is $\gamma _{i}$. $\hat{\sigma}%
_{x,z}^{i}$ and $\alpha$ are the Pauli operator of  the $i$th particle and coupling angle
between the spin and bath, respectively. For simplicity, in the numerical simulation of the master equation and the calculation of the QFI, we assume that the coupling angle and decay rate are uniform for all particles.

For the case of $N=2$, as depicted in Fig.~\ref{fig1s}(a), the introduction of NPC noise notably boosts the estimation precision of $B$. Fig.~\ref{fig1s}(b) illustrates that the optimal coupling angle remains $\alpha=k \pi/4$ ($k$ being an odd number). 
Furthermore, when estimating the parameter $\vartheta$, the maximum value of $F_\vartheta$ without noise is $16$ for $N=2$. Fig.~\ref{fig1s}(c) demonstrates that this value can be exceeded with the assistance of NPC noise. These findings align with the single-particle case presented in Figs.~1 and 2 of the main text.

\begin{figure}[hbt!]
	\centering
	\includegraphics[width=0.5\linewidth]{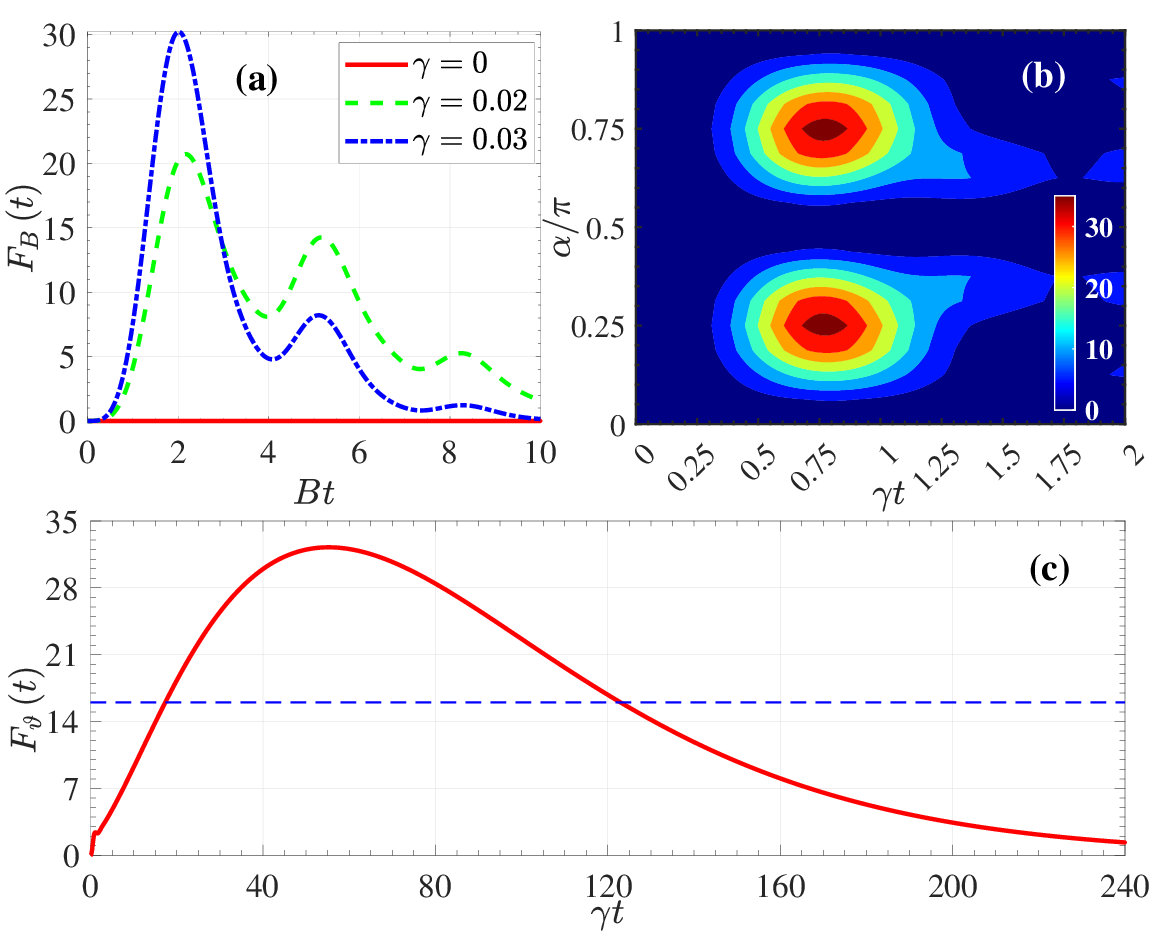}
	\caption{(a) $F_{B}$ of a two particle system versus encoding time $t$ with various decay rate, where $\alpha =\pi /4$ and $\vartheta =0$. (b) The density plot of $F_{B}$ as a function of $\alpha $ and $\gamma t$, where $\vartheta =0$. (c) $F_{\vartheta}$ versus encoding time, where $\alpha=0.3\pi$ and $\vartheta =\pi/3$. The blue dashed line denotes the noise-free maximum $F_\vartheta$. In all subgraphs, $B=0.1$ (used as a scale) and the initial state $\left\vert \Phi _{\text{tot}}\left( 0\right) \right\rangle =\left[( \left\vert e\right\rangle +\left\vert g\right\rangle)/\sqrt{2}\right]^{\otimes 2} $.}
	\label{fig1s}
\end{figure}   

In Fig.~3 of the main text, we demonstrated the noise-assisted surpassing of the maximum $F_\vartheta$ for $N=2,3,4$ when the initial state of the system was assumed to be a non-entangled state. Herein, we have modified the initial state to the GHZ state (maximally entangled state) given by $\left\vert \Phi _{\text{tot}}\left( 0\right) \right\rangle =( \left\vert e\right\rangle ^{\otimes N}+\left\vert g\right\rangle^{\otimes N})/\sqrt{2} $. As evident from Fig.~\ref{fig2s}, the conclusions obtained are similar to those for the non-entangled initial state. However, interestingly, we observe that under identical conditions, the use of the GHZ state is more effective in surpassing the precision limit without noise compared to the non-entangled initial state. This advantage is reflected in a more lenient selection of $\alpha$ and a reduction in encoding time. For instance, when $N=4$ and $\alpha=0.3\pi$, compare the blue dotted line in Fig.~3(d) of the main text with the purple dotted-dashed line in Fig.~\ref{fig2s}(c).

\begin{figure}[hbt!]
	\centering
	\includegraphics[width=0.55\linewidth]{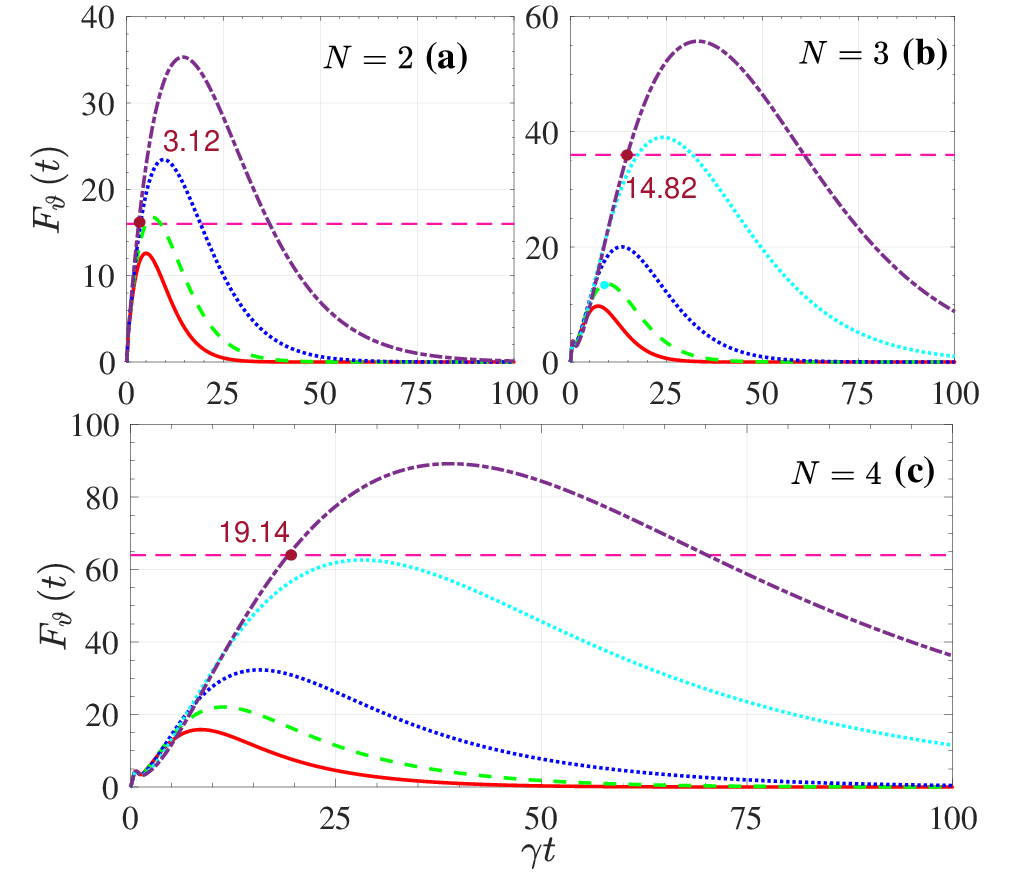}
	\caption{The QFI versus the  $\gamma t$ for different coupling angles $\alpha$, where the initial state of the system is $\left\vert \Phi _{\text{tot}}\left( 0\right) \right\rangle =( \left\vert e\right\rangle ^{\otimes N}+\left\vert g\right\rangle
		^{\otimes N})/\sqrt{2} $ and the estimated parameter $\vartheta=\pi/3$. (a) The particle number $N=2$, and the red solid line, green dashed line, blue dotted line, and purple dotted-dashed line correspond to coupling angle $\alpha=0.26\pi, 0.27\pi, 0.28\pi, 0.29$ in sequence, respectively. (b,c) The particle number $N=2, 3$. The red solid line, green dashed line, blue dotted line, sky blue dotted line, and purple dotted-dashed line correspond to coupling angle $\alpha=0.26\pi, 0.27\pi, 0.28\pi, 0.294\pi,0.3\pi$ in sequence, respectively. The pink dashed lines in each subgraph correspond to the maximum QFI by coherent dynamics, and the brown numbers represent the time required to surpass them. }
	\label{fig2s}
\end{figure}

\section{VIII. The condition for taking an upper bound on channel QFI}
Now, let's discuss the inequality recently proposed by Wang et al \cite{CSWXG},
\begin{equation}
\label{Ews}
F_{\lambda }^{\left( c\right) }\left( t\right) \leq \left[
\int\nolimits_{0}^{t}|| \partial \hat{H}_{\lambda }\left( s\right)
/\partial \lambda || ds\right] ^{2},
\end{equation}
where $\hat{H}_{\lambda
}\left( t\right) $ is the parameter $\lambda$-dependent Hamiltonian of the system, $||\hat{A}||=A_{\text{max}}-A_{\text{min}}$, with $A_{\text{max}}$ $(A_{\text{min}})$ the maximum (minimum) eigenvalue of operator $\hat{A}$, and $F_{\lambda }^{\left( c\right)}\left( t\right) $ is the maximum QFI achievable by optimizing initial state. Here, we point out that in many cases, $F_{\lambda }^{\left( c\right) }\left( t\right)$ cannot reach this upper bound. 
This opens avenues for enhancing $F_{\lambda }^{\left( c\right) }\left( t\right)$ through quantum control, auxiliary systems, and even noise utilization. Consequently, identifying the conditions for equality in inequality (\ref{Ews}) has become crucial. Given the reliance on the triangle inequality of seminorm, $||\hat{A}+\hat{B}||\leq ||\hat{A}||+||\hat{B}||$, in deriving (\ref{Ews}), we initially establish its equality conditions.

Let's define $\hat{L}=\hat{A}+\hat{B}$. $L_{\text{max}}$ and $L_{\text{min}}$ represent the maximum and minimum eigenvalues of $\hat{L}$, with the corresponding eigenstates $|L_{\text{max}}\rangle $ and $|L_{\text{min}}\rangle$. Utilizing these definitions, we can derive
\begin{eqnarray}
\label{Eqr78}
\langle L_{\text{max}}|\hat{L}|L_{\text{max}}\rangle  &=&\langle L_{\text{max}}|\hat{A}|L_{\text{max}}\rangle +\langle L_{%
	\text{max}}|\hat{B}|L_{\text{max}}\rangle   \notag \\
&\leq &A_{\text{max}}+B_{\text{max}}, \\
\langle L_{\text{min}}|\hat{L}|L_{\text{min}}\rangle  &=&\langle 
L_{\text{min}}|\hat{A}|L_{\text{min}}\rangle +\langle L_{%
	\text{min}}|\hat{B}| L_{\text{min}}\rangle   \notag \\
&\geq &A_{\text{min}}+B_{\text{min}}. \label{Eqr79}
\end{eqnarray}%
Consequently, we arrive at the inequality
\begin{eqnarray}
||\hat{A}+\hat{B}||&=& \langle L_{\text{max}}|\hat{L}|L_{\text{max}}\rangle -\langle 
L_{\text{min}}|\hat{L}|L_{\text{min}}\rangle   \notag \\
&\leq &(A_{\text{max}}+B_{\text{max}})-(A_{\text{min}}+%
B_{\text{min}})  \notag \\
&=&||\hat{A}||+||\hat{B}||.
\end{eqnarray} 
Based on Eqs.~(\ref{Eqr78}) and (\ref{Eqr79}), the requirement for the inequality to hold with equality is that $\hat{A}$ and $\hat{B}$ share a common set of eigenstates, indicating $[\hat{A},\hat{B}]=0$. Extending this inequality to include more operators is straightforward, for instance,
\begin{equation}
\label{Eqr78B}
||\hat{A}+\hat{B}+\hat{C}||\leq ||\hat{A}||+||\hat{B}||+||\hat{C}||,
\end{equation}
with the equality sign achieved when all three operators mutually commute with each other.

For closed systems, the transformed local generator is~\cite{pang2017optimal} 
\begin{equation}
\hat{h}_{\lambda }\left( t\right) =\int_0^{t}\hat{\Pi}_{\lambda }\left( s\right)ds,
\end{equation}%
where 
\begin{equation}
\hat{\Pi}_{\lambda }\left( s\right)
=\hat{U}_{\lambda }^{\dag }(0\rightarrow s)\tfrac{\partial \hat{H}_{\lambda}\left( s\right) }{\partial \lambda }\hat{U}_{\lambda }(0\rightarrow s).
\label{S75}
\end{equation}
Therefore, the channel QFI satisfies the following inequality~\cite{CSWXG,PRLKK},
\begin{eqnarray}
\label{Eq763}
F_{\lambda }^{\left( c\right) }\left( t\right)  &=&\left\Vert \hat{h}%
_{\lambda }\left( t\right) \right\Vert =\left\Vert \int\nolimits_{0}^{t}\hat{%
	\Pi}_{\lambda }\left( s\right) ds\right\Vert   \notag \\
&=&\left\Vert \lim_{n\rightarrow \infty }\sum\limits_{k=1}^{n}\hat{\Pi}%
_{\lambda }\left( \xi _{k}\right) \frac{t}{n}\right\Vert   \notag \\
&\leq &\lim_{n\rightarrow \infty }\sum\limits_{k=1}^{n}\left\Vert \hat{\Pi}%
_{\lambda }\left( \xi _{k}\right) \right\Vert \frac{t}{n}  \notag \\
&=&\int\nolimits_{0}^{t}\left\Vert \hat{\Pi}_{\lambda }\left( s\right)
\right\Vert ds  \notag \\
&=&\int\nolimits_{0}^{t}\left\Vert \hat{U}_{\lambda }^{\dag }(0\rightarrow s)%
\tfrac{\partial \hat{H}_{\lambda }\left( s\right) }{\partial \lambda }\hat{U}%
_{\lambda }(0\rightarrow s)\right\Vert ds  \notag \\
&=&\int\nolimits_{0}^{t}\left\Vert \tfrac{\partial \hat{H}_{\lambda }\left(
	s\right) }{\partial \lambda }\right\Vert ds,
\end{eqnarray}%
where $\xi _{k}\in \left[ \frac{(k-1)t}{n},\frac{kt}{n}\right] (k=1,2,3....n)$.
Because $n\rightarrow \infty $ $\left( \frac{t}{n}\rightarrow 0\right) $,
combined with Eq.~(\ref{Eqr78B}), the condition for taking the equal sign of the
above equation is $\left[ \hat{\Pi}_{\lambda }\left( t_{i}\right) ,\hat{\Pi}_{\lambda }\left(
t_{j}\right) \right] \equiv 0$ $\left( 0\leq t_{i},t_{j}\leq t\right) $.
According to the definition of $\hat{\Pi}_{\lambda }$ given in Eq. (\ref{S75}), it is evident that when the estimated parameter $\lambda$ is an overall factor of the Hamiltonian $\hat{H}_{\lambda}$, the equal sign can be achieved. Consequently, the upper bound of $ F_{\lambda }^{\left( c\right) }\left( t\right)$ established by inequality (\ref{Ews}) can be achieved. In this case, the ultimate precision limit of Hermitian sensors cannot be enhanced through the addition of auxiliary parameter-independent Hamiltonian extensions or noises. Conversely, if the estimated parameter is not an overall factor of the Hamiltonian, $ F_{\lambda }^{\left( c\right) }\left( t\right)$ does not reach the upper limit established by inequality (\ref{Ews}), indicating potential for further optimization.

\end{widetext}
	
\end{document}